\documentclass[]{aa}  

\usepackage{natbib}
\bibpunct{(}{)}{;}{a}{}{,} 
\setcitestyle{open={(},author,year,close={)}}
\usepackage{graphicx}
\usepackage{txfonts}

\begin{document} 

   \title{Retrieving stellar parameters and dynamics of AGB stars \\ 
   with \textit{Gaia} parallax measurements and CO$^5$BOLD RHD simulations}
   \author{E. Béguin
          \inst{1}
          \and
          A. Chiavassa\inst{1}
          \and
          A. Ahmad\inst{2}
          \and
          B. Freytag\inst{2}
          \and
          S. Uttenthaler\inst{3}
          }

   \institute{Universit\'e C\^ote d'Azur, Observatoire de la C\^ote d'Azur, CNRS, Lagrange, CS 34229, Nice,  France \\
              \email{elysabeth.beguin@oca.eu}
            \and
            Theoretical Astrophysics, Department of Physics and Astronomy, Uppsala University, Box~516, SE-751~20 Uppsala, Sweden
            \and
            Institute of Applied Physics, TU Wien, Wiedner Hauptstra\ss e 8-10, 1040 Vienna, Austria}

   \date{Received April 4, 2024; accepted July 16, 2024}

   \abstract
   {The complex dynamics of asymptotic giant branch (AGB) stars and the resulting stellar winds have a significant impact on the measurements of stellar parameters and amplify their uncertainties. Three-dimensional (3D) radiative hydrodynamic (RHD) simulations of convection suggest that convection-related structures at the surface of AGB star affect the photocentre displacement and the parallax uncertainty measured by \textit{Gaia}.}
   {We explore the impact of the convection on the photocentre variability and aim to establish analytical laws between the photocentre displacement and stellar parameters to retrieve such parameters from the parallax uncertainty.}
   {We used a selection of $31$ RHD simulations with CO$^5$BOLD and the post-processing radiative transfer code O\textsc{ptim3D} to compute intensity maps in the \textit{Gaia} G band [320--1050 nm]. From these maps, we calculated the photocentre position and temporal fluctuations. We then compared the synthetic standard deviation to the parallax uncertainty of a sample of $53$ Mira stars observed with \textit{Gaia}.}
   {The simulations show a displacement of the photocentre across the surface ranging from $4$ to $13 \%$ of the corresponding stellar radius, in agreement with previous studies. We provide an analytical law relating the pulsation period of the simulations and the photocentre displacement as well as the pulsation period and stellar parameters. By combining these laws, we retrieve the surface gravity, the effective temperature, and the radius for the stars in our sample.}
   {Our analysis highlights an original procedure to retrieve stellar parameters by using both state-of-the-art 3D numerical simulations of AGB stellar convection and parallax observations of AGB stars. This will help us refine our understanding of these giants.}


   \titlerunning{\textit{Gaia} parallax uncertainties and RHD simulations}

   \maketitle
%
 
\section{Introduction}

Low- to intermediate-mass stars ($0.8-8\, \mathrm{M_\odot}$) evolve into the asymptotic giant branch (AGB), in which they undergo complex dynamics characterised by several processes including convection, pulsations, and shockwaves. These processes trigger strong stellar winds \citep[$10^{-8}$--$10^{-5}\, \mathrm{M_\odot/yr}$, ][]{de_beck_probing_2010} that significantly enrich the interstellar medium with various chemical elements \citep{hofner_mass_2018}. These processes and stellar winds also amplify uncertainties of stellar parameter determinations with spectro-photometric techniques, like the effective temperature, which in turn impacts the determination of mass-loss rates \citep{hofner_mass_2018}. In particular, Mira stars are peculiar AGB stars, showing extreme magnitude variability (larger than $2.5$ mag in the visible) due to pulsations over periods of $100$ to $1000$ days \citep{decin_evolution_2021}.

In \cite{chiavassa_pasquato_2011, chiavassa_heading_2018, chiavassa_probing_2022}, 3D radiative hydrodynamics (RHD) simulations of convection computed with CO$^5$BOLD \citep{freytag_simulations_2012, freytag_advances_2013, freytag_boundary_2017} reveal the AGB photosphere morphology to be made of a few large-scale, long-lived convective cells and some short-lived and small-scale structures that cause temporal fluctuations on the emerging intensity in the \textit{Gaia} G band [320--1050 nm]. The authors suggest that the temporal convective-related photocentre variability should substantially impact the photometric measurements of \textit{Gaia}, and thus the parallax uncertainty. In this work, we use $31$ recent simulations to establish analytical laws between the photocentre displacement and the pulsation period and then between the pulsation period and stellar parameters. We combine these laws and apply them to a sample of $53$ Mira stars from \cite{uttenthaler_interplay_2019} to retrieve their effective stellar gravity, effective temperature, and radius thanks to their parallax uncertainty from \textit{Gaia} Data Release 3\footnote{GDR3 website: https://www.cosmos.esa.int/web/gaia/data-release-3} (GDR3) \citep{gaia_collaboration_gaia_2016, gaia_collaboration_gaia_2023}.

\section{Overview of the radiative hydrodynamics simulations}

In this section, we present the simulations and the theoretical relations between the stellar parameters and the pulsation period. We also present how we compute the standard deviation of the photocentre displacement and its correlation with the pulsation period.

\subsection{Methods}

We used RHD simulations of AGB stars computed with the code CO$^5$BOLD \citep{freytag_simulations_2012,freytag_advances_2013,freytag_boundary_2017}. It solves the coupled non-linear equations of compressible hydrodynamics and non-local radiative energy transfer, assuming solar abundances, which is appropriate for M-type AGB stars. The configuration is ‘star-in-a-box’, which takes into account the dynamics of the outer convective envelope and the inner atmosphere. Convection and pulsations in the stellar interior trigger shocks in the outer atmosphere, giving a direct insight into the stellar stratification. Material can levitate towards layers where it can condensate into dust grains \citep{freytag_global_2023}. However, models used in this work do not include dust. 


We then post-processed a set of temporal snapshots from the RHD simulations using the radiative transfer code O\textsc{ptim3D} \citep{chiavassa_radiative_2009}, which takes into account the Doppler shifts, partly due to convection, in order to compute intensity maps integrated over the \textit{Gaia} G band [320--1050 nm]. The radiative transfer is computed using pre-tabulated extinction coefficients from MARCS models \citep{gustafsson_grid_2008} and solar abundance tables \citep{asplund_chemical_2009}. 

\subsection{Characterising the AGB stellar grid}

We used a selection of simulations from \cite{freytag_global_2017}, \cite{chiavassa_heading_2018} [abbreviation: F17+C18], \cite{ahmad_properties_2023} [abbreviation: A23] and some new models [abbreviation: This work], in order to cover the $2000$--$10000\ \mathrm{L_\odot}$ range. The updated simulation parameters are reported in Table \ref{tab_simus}. 

In particular, $24$ simulations have a stellar mass equal to $\mathrm{1.0\, M_\odot}$, and seven simulations have one equal to $\mathrm{1.5\, M_\odot}$. In the rest of this work, we denote by a $1.0$ subscript the laws or the results obtained from the analysis of the $\mathrm{1.0\, M_\odot}$ simulations; and by a $1.5$ subscript those obtained from the analysis of the $\mathrm{1.5\, M_\odot}$ simulations. 

\begin{figure}[h!]
    \centering
    \includegraphics[width=\columnwidth]{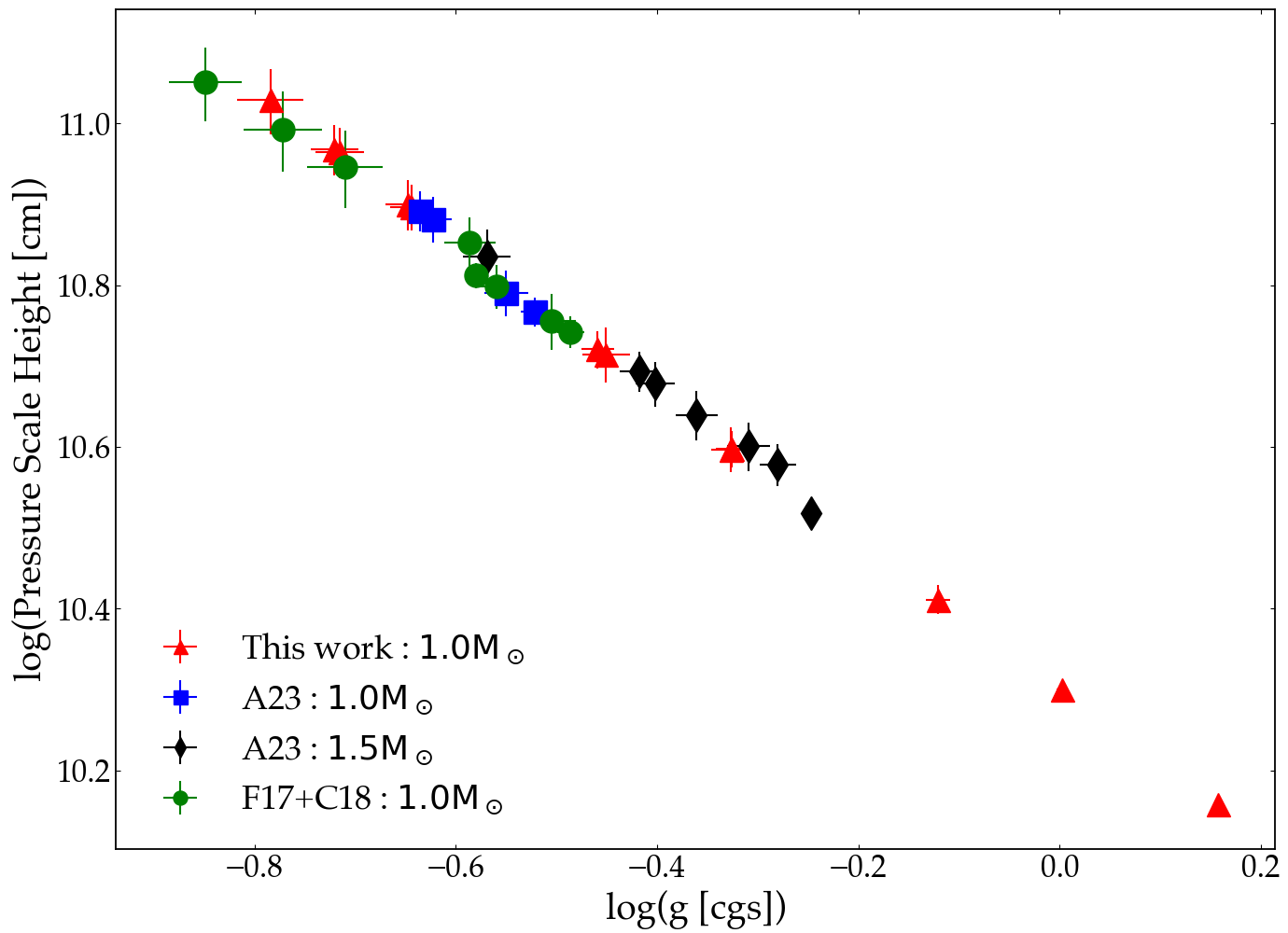}
    \caption{Log-log plot of the surface gravity [cgs] versus the pressure scale height [cm]. As the effective temperature is near constant among all our models, we expect a linear relation.}
    \label{fig_pressure_scale_gravity}
\end{figure}

The pressure scale height is defined as $\mathrm{H_p}\, = \,\frac{k_B \mathrm{T_{eff}}}{\mu g}$, with $k_B$ the Boltzmann constant, $\mathrm{T_{eff}}$ the effective surface temperature, $\mu$ the mean molecular mass, and $g$ the local surface gravity. The lower the surface gravity is, the larger the pressure scale height becomes, and so the larger the convective cells can grow (see Fig. \ref{fig_pressure_scale_gravity} and \cite{freytag_global_2017}). 


Moreover, \cite{freytag_scale_1997} estimated that the characteristic granule size scales linearly with $H_P$. Interplay between large-scale convection and radial pulsations results in the formation of giant and bright convective cells at the surface. The resulting intensity asymmetries directly cause temporal and spatial fluctuations of the photocentre. The larger the cells, the larger the photocentre displacement. This results in a linear relation between the photocentre displacement and the pressure scale height \citep{chiavassa_pasquato_2011,chiavassa_heading_2018}.
However, \cite{chiavassa_pasquato_2011} found it is no longer true for $\mathrm{H_p}$ greater than $2.24 \times 10^{10} \mathrm{cm}$ for both interferometric observations of red supergiant stars and 3D simulations, suggesting that the relation for evolved stars is more complex and depends on $\mathrm{H_p}$ (Fig. 18 in the aforementioned article).

Concerning the pulsation period, \cite{ahmad_properties_2023} performed a fast Fourier transform on spherically averaged mass flows of the CO$^5$BOLD snapshots to derive the radial pulsation periods. The derived bolometric luminosity--period relation suggests good agreement between the pulsation periods obtained from the RHD simulations and from available observations. They are reported in columns 9 and 10 in Table \ref{tab_simus}. 

\begin{figure}[h!]
    \centering
    \includegraphics[width=\columnwidth]{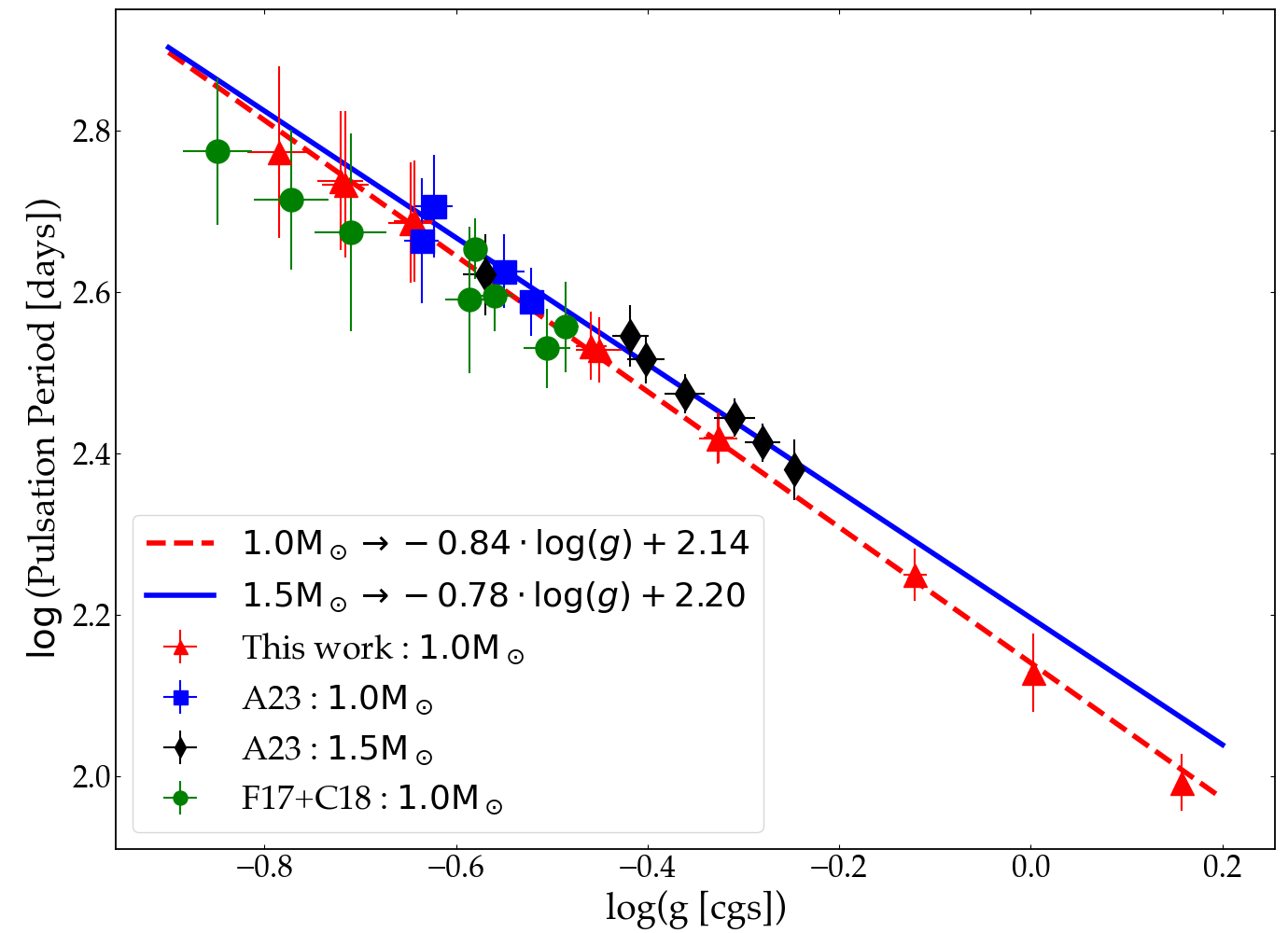}
    \caption{Log-log plot of the pulsation period [days] versus the surface gravity, $\mathrm{g}$ [cgs], which follows a linear law whose parameters were computed with a non-linear least-squares method, given in Eqs. (\ref{formule_logg_logPpuls10}) and (\ref{formule_logg_logPpuls15}).}
    \label{fig_logg_ppsul}
\end{figure}

\cite{ahmad_properties_2023} found a correlation between the pulsation period and the surface gravity ($\mathrm{g} \propto \mathrm{M_\star}/\mathrm{R_\star^2}$, see Eq. (4) in the aforementioned article). Thus, the pulsation period increases when the surface gravity decreases. In agreement with this study, we found a linear relation between $\log(\mathrm{P_{puls}})$ and $\mathrm{log(g)}$, see Fig. \ref{fig_logg_ppsul}. By minimising the sum of the squares of the residuals between the data and a linear function as in the non-linear least-squares problem, we computed the most suitable parameters of the linear law. We also computed the reduced $\bar{\chi}^{2}$ for the $\mathrm{1.0\, M_\odot}$ simulations and for the $\mathrm{1.5\, M_\odot}$ simulations: $\bar{\chi}_{1.0}^{2}\, = \,2.2$ and $\bar{\chi}_{1.5}^{2}\, = \,0.4$. The linear law found in each case is expressed as follows:  

\begin{equation}
    \log(\mathrm{P_{puls}})\, = \,-0.84 \cdot \log(\mathrm{g}) + 2.14\mathrm{,\ \ for\ M_\star\, = \,1.0\, M_\odot}
    \label{formule_logg_logPpuls10}
\end{equation}

\begin{equation}
    \log(\mathrm{P_{puls}})\, = \,-0.78 \cdot \log(\mathrm{g}) + 2.20\mathrm{,\ \ for\ M_\star\, = \,1.5\, M_\odot}
    \label{formule_logg_logPpuls15}
.\end{equation}

It is important to note that the $\bar{\chi}_{1.5}^{2}$ value is lower than $\bar{\chi}_{1.0}^{2}$ because there are only seven points to fit (i.e. seven RHD simulations at $\mathrm{1.5\, M_\odot}$) and they are less scattered.

\begin{figure}[h!]
    \centering
    \includegraphics[width=\columnwidth]{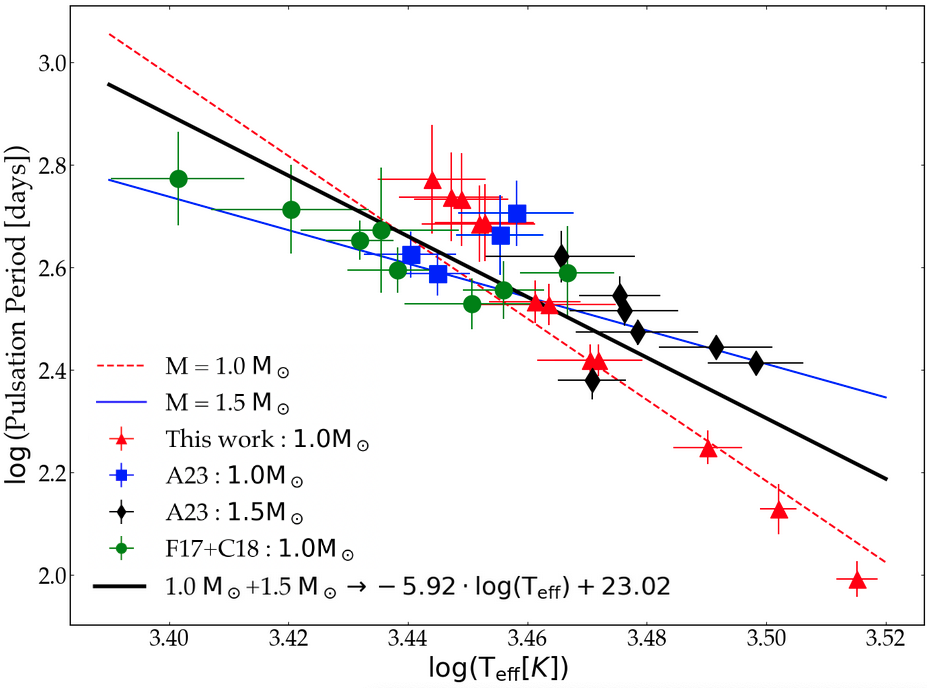}
    \caption{Log-log plot of the pulsation period [days] versus the effective temperature, $\mathrm{T_{eff}}$ [K], which is in agreement with photosphere dynamics. We notice two groups of data (above and below the black curve) that are not linked to mass. The linear relation is given in Eq. (\ref{formule_logT_logPpuls}).}
    \label{fig_Ppuls_T}
\end{figure}

We compared the pulsation period, $\mathrm{P_{puls}}$, with the effective temperature, $\mathrm{T_{eff}}$. \cite{ahmad_properties_2023} showed that the pulsation period decreases when the temperature increases. A linear correlation is confirmed with our simulations, as is displayed in Fig. \ref{fig_Ppuls_T}. However, we do not see any clear differentiation between the law found from the $\mathrm{1.0\, M_\odot}$ simulations and the law from the $\mathrm{1.5\, M_\odot}$ simulations so we chose to use all simulations to infer a law between $\mathrm{P_{puls}}$ and $\mathrm{T_{eff}}$ (Eq. (\ref{formule_logT_logPpuls}) and the black curve in Fig. \ref{fig_Ppuls_T}). We computed the reduced $\bar{\chi}^2\, = \,77$ and the parameters of the linear law are expressed as follows: 

\begin{equation}
    \log(\mathrm{P_{puls}})\, = \,-5.92 \cdot \log(\mathrm{T_{eff}}) + 23.02
    \label{formule_logT_logPpuls}
.\end{equation}

\begin{figure}[h!]
    \centering
    \includegraphics[width=\columnwidth]{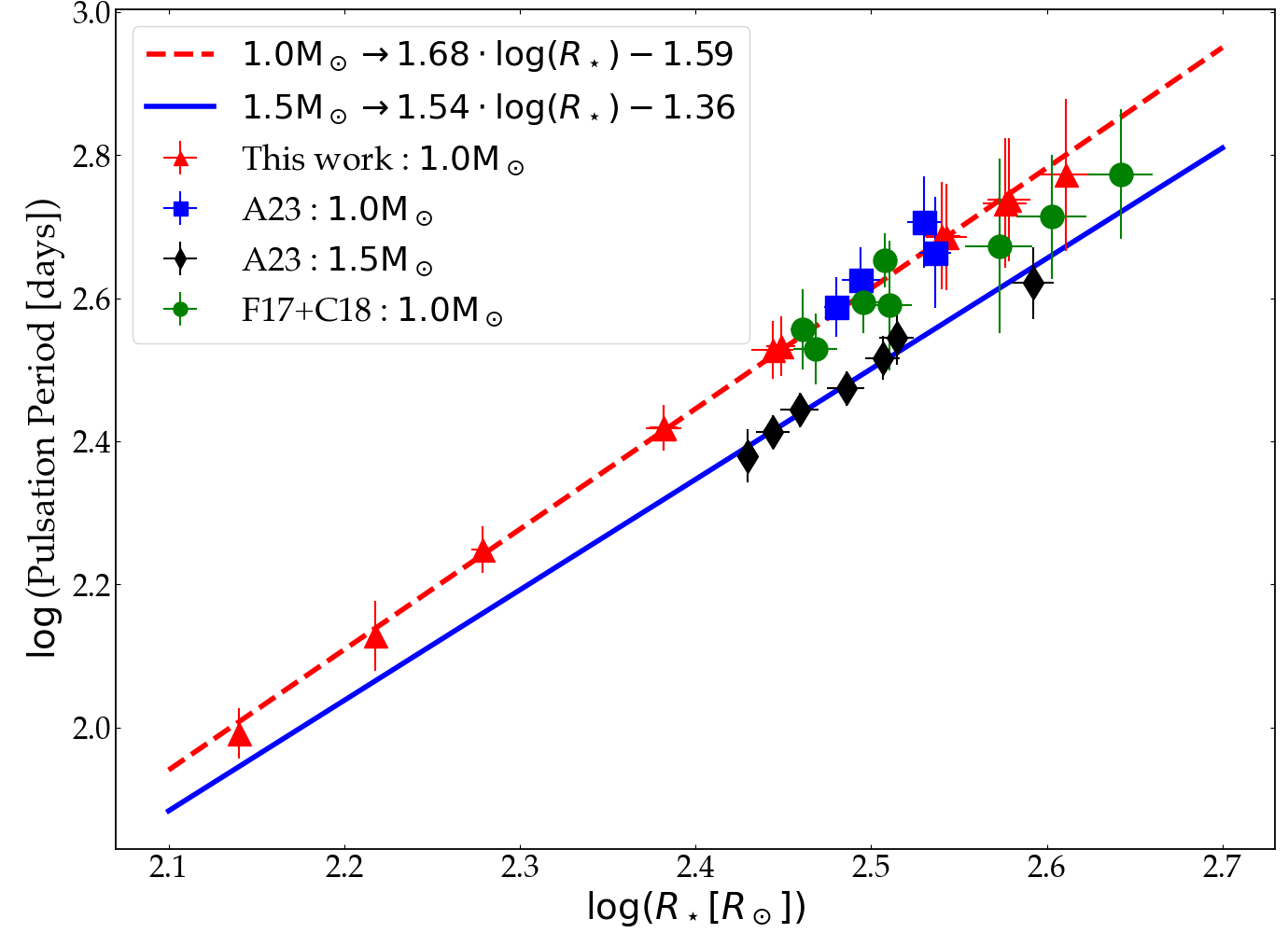}
    \caption{Log-log of the pulsation period [days] versus the stellar radius $\mathrm{R_\star}$[$\mathrm{R_\odot}$], which is in agreement with photosphere dynamics. The linear laws' expressions are given in Eqs. (\ref{formule_logR_logPpuls_10M}) and (\ref{formule_logR_logPpuls_15M}).}
    \label{fig_logR_logPpuls}
\end{figure}

\cite{ahmad_properties_2023} found a correlation between the pulsation period and the inverse square root of the stellar mean-density ($\mathrm{M_\star}/\mathrm{R_\star^3}$, see Eq. (4) in the aforementioned article). In agreement with this study, we found a linear correlation between $\log(\mathrm{P_{puls}})$ and $\mathrm{log(R_\star)}$, with $\mathrm{R_\star}$ the stellar radius (Fig. \ref{fig_logR_logPpuls}). We computed with the least-squares method $\bar{\chi}_{1.0}^{2}\, = \,2.0$ and $\bar{\chi}_{1.5}^{2}\, = \,0.4$ and the parameters of the linear law are expressed as follows:

\begin{equation}
    \log(\mathrm{P_{puls}})\, = \,1.68 \cdot \log(\mathrm{R_\star}) - 1.59\mathrm{,\ \ for\ M_\star\, = \,1.0\, M_\odot}
    \label{formule_logR_logPpuls_10M}
\end{equation}

\begin{equation}
    \log(\mathrm{P_{puls}})\, = \,1.54 \cdot \log(\mathrm{R_\star}) - 1.36\mathrm{,\ \ for\ M_\star\, = \,1.5\, M_\odot}
    \label{formule_logR_logPpuls_15M}
.\end{equation}

\begin{figure}[h!]
    \centering
    \includegraphics[width=\columnwidth]{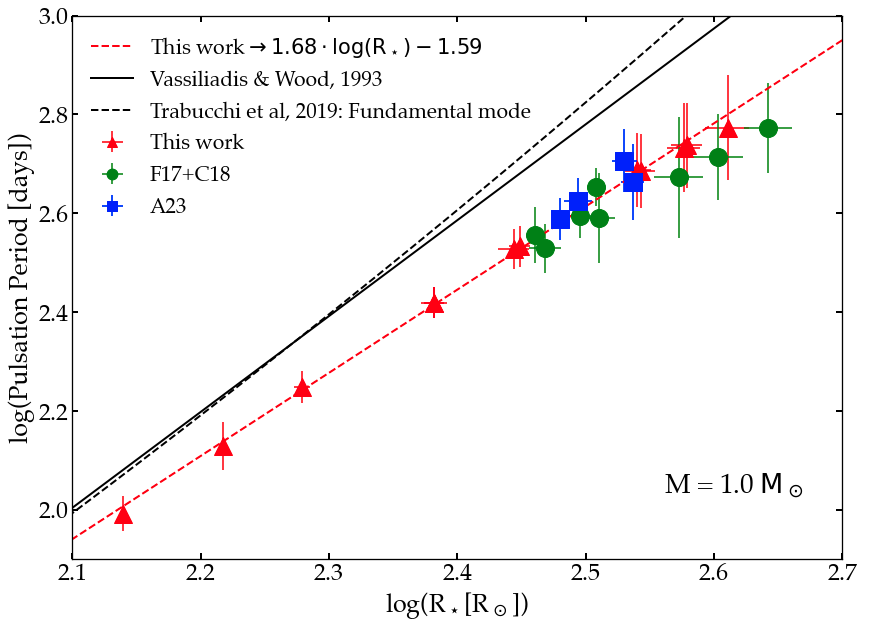}
    \caption{\textbf{Case of the $\mathrm{1.0\, M_\odot}$ simulations}: Log-log of the pulsation period [days] versus the stellar radius, $\mathrm{R_\star}$[$\mathrm{R_\odot}$]. We compare the analytical law established here (dashed red curve) with the law from \cite{vassiliadis_evolution_1993} (black curve) and with the fundamental mode from \cite{trabucchi_modelling_2019} (dashed black curve). Overall, we found similar trends, only separated by an offset. The pulsation periods of the simulations are shorter than expected.}
    \label{VassTrab10}
\end{figure}

\begin{figure}[h!]
    \centering
    \includegraphics[width=\columnwidth]{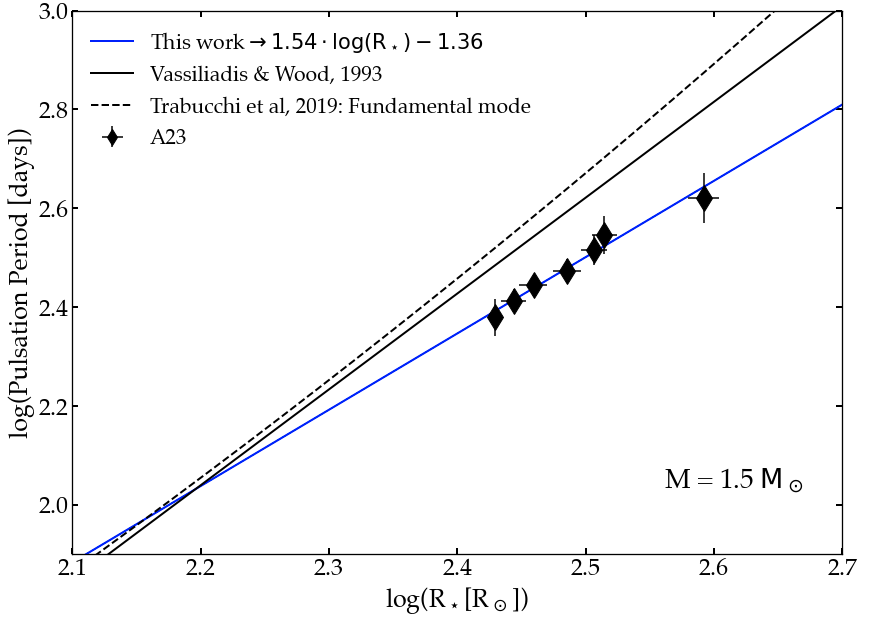}
    \caption{\textbf{Case of the $\mathrm{1.5\, M_\odot}$ simulations}: Same as in Fig. \ref{VassTrab10}. In this case, the mass appears in the equations from \cite{trabucchi_modelling_2019} as its logarithm does not equal zero anymore.}
    \label{VassTrab15}
\end{figure}

We compared our results with the relation found by \cite{vassiliadis_evolution_1993} — $\log(\mathrm{P_{puls}})\, =\, -2.07 + 1.94 \log(\mathrm{R_\star}/\mathrm{R_\odot}) - 0.9 \log(\mathrm{M_\star}/\mathrm{M_\odot})$ — and with the fundamental mode of long-period variables, Eq. (12), found by \cite{trabucchi_modelling_2019} (see Figs. \ref{VassTrab10} and \ref{VassTrab15}). We used the solar metallicity and helium mass function from \cite{asplund_chemical_2009} and the reference carbon-to-oxygen ratio from \cite{trabucchi_modelling_2019}. Overall, we have the same trends, but we notice that the laws we obtained are less steep than the relation from \cite{vassiliadis_evolution_1993} and the fundamental mode from \cite{trabucchi_modelling_2019}, meaning that the pulsation periods of our simulations are shorter than expected.

\subsection{Photocentre variability of the RHD simulations in the Gaia G band}

\begin{figure*}[ht]
    \centering
    \includegraphics[width=\textwidth]{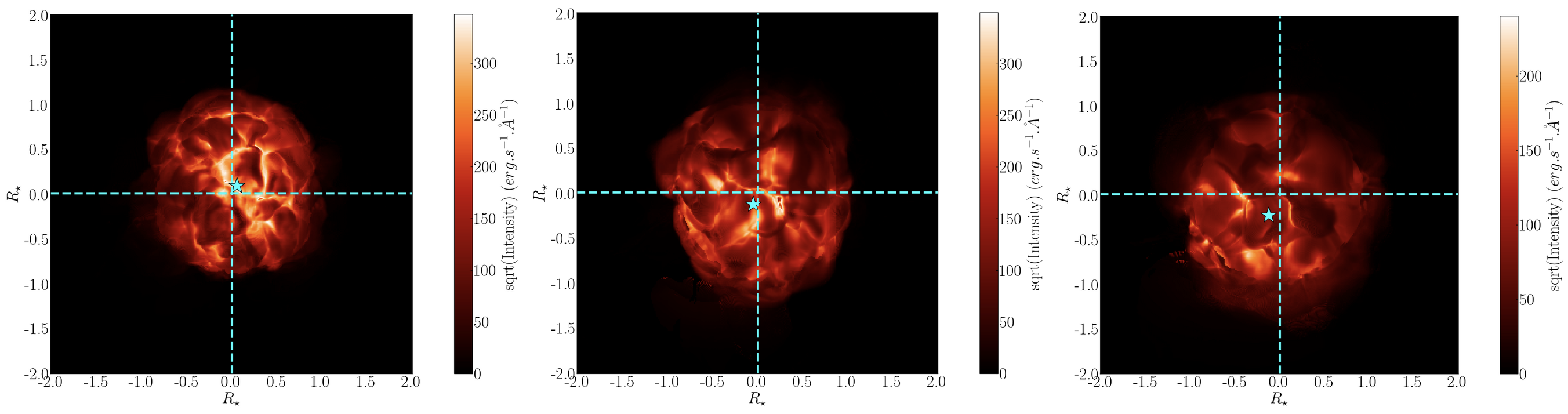}
    \caption{Temporal evolution of an AGB simulation. The intensity maps are in $\mathrm{erg \cdot s^{-1}} \cdot$ Å$^{-1}$. The star indicates the position of the photocentre at the given time for the simulation \textit{st28gm05n028}. The dashed lines intersect at the geometric centre of the image.}
    \label{fig_frise}
\end{figure*}

For each intensity map computed (for example, Fig. \ref{fig_frise}), we calculated the position of the photocentre as the intensity-weighted mean of the x-y positions of all emitting points tiling the visible stellar surface according to

\begin{equation}
    P_x\, = \,\frac{\sum ^N _{i=1} \sum ^N _{j=1} I(i,j) \cdot x(i,j)}{\sum ^N _{i=1} \sum ^N _{j=1} I(i,j)}
\end{equation}

\begin{equation}
    P_y\, = \,\frac{\sum ^N _{i=1} \sum ^N _{j=1} I(i,j) \cdot y(i,j)}{\sum ^N _{i=1} \sum ^N _{j=1} I(i,j)}
,\end{equation}

where $I(i,j)$ is the emerging intensity for the grid point $(i,j)$ with co-ordinates $x(i,j),y(i,j)$ and $N$ the number of points in each co-ordinate of the simulated box.

Large-scale convective cells drag hot plasma from the core towards the surface, where it cools down and sinks \citep{freytag_global_2017}. Coupled with pulsations, this causes optical depth and brightness temporal and spatial variability, moving the photocentre position \citep{chiavassa_pasquato_2011,chiavassa_heading_2018}. Thus, in the presence of brightness asymmetries, the photocentre will not coincide with the barycentre of the star. Figure \ref{fig_frise} displays the time variability of the photocentre position (blue star) for three snapshots of the simulation \textit{st28gm05n028}. The dashed lines intersect at the geometric centre of the image.

We computed the time-averaged photocentre position, $\langle P_x \rangle$ and $\langle P_y \rangle$, for each Cartesian co-ordinate in astronomical units, [$\mathrm{AU}$], the time-averaged radial photocentre position, $\langle P \rangle$ as $\langle P \rangle\, = \,(\langle P_x \rangle ^2+\langle P_y \rangle^2)^{1/2}$, and its standard deviation, $\mathrm{\sigma_P}$, in $\mathrm{AU}$ and as a percentage of the corresponding stellar radius, $\mathrm{R_\star}$ [\% of $R_\star$] (see Table \ref{tab_simus}). 

\begin{figure}[ht]
    \centering   \includegraphics[width=\columnwidth]{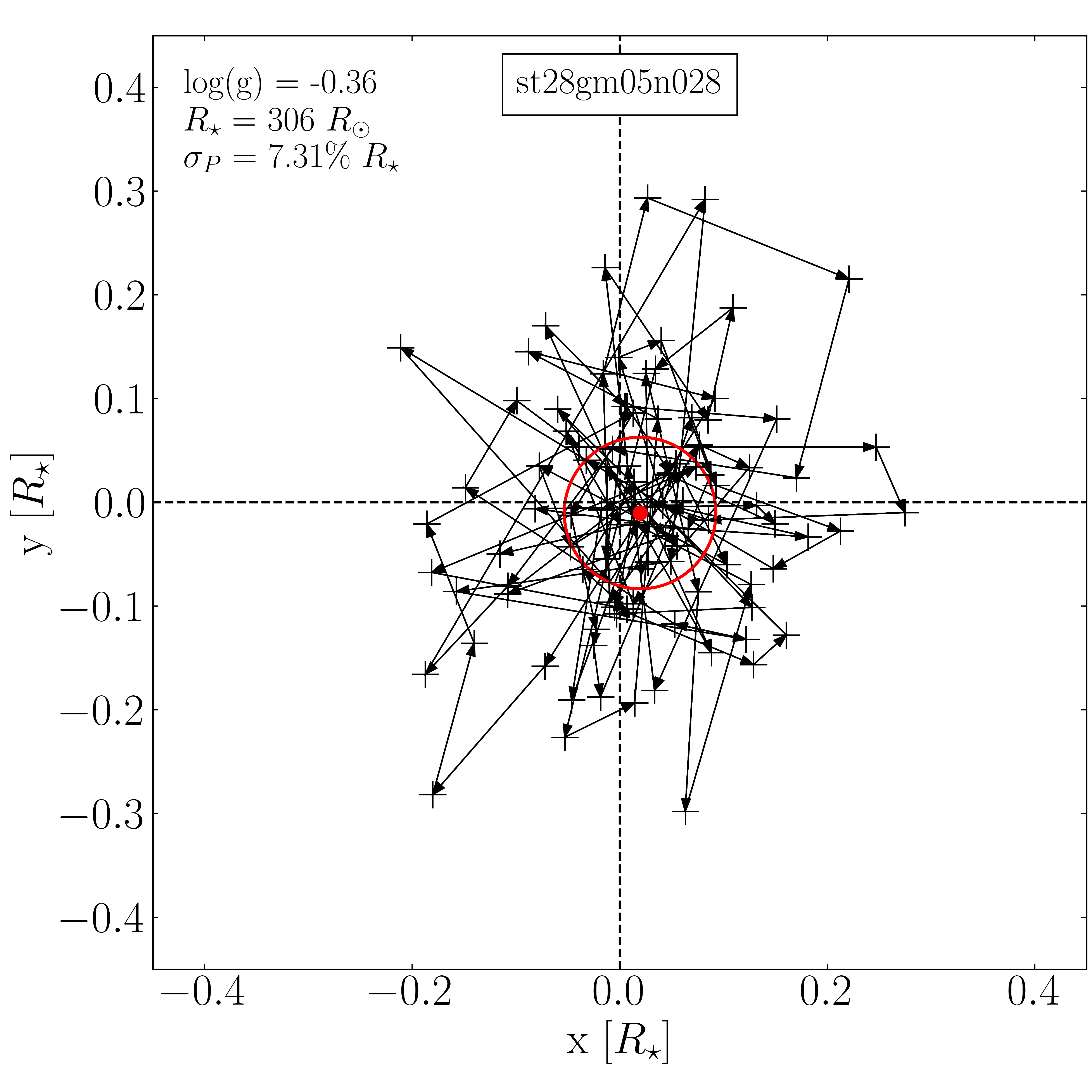}
    \caption{Temporal evolution of the photocentre displacement for an AGB simulation, here \textit{st28gm05n028}, same as in Fig. \ref{fig_frise}. The dashed lines intersect at the geometric centre of the image. The red dot indicates the average position of the photocentre and the red circle its standard deviation.}
    \label{displacement_plots}
\end{figure}

Figure \ref{displacement_plots} displays the photocentre displacement over the total duration for the simulation \textit{st28gm05n028}, with the average position as the red dot and $\mathrm{\sigma_P}$ as the red circle radius (see additional simulations in Figs. B1 to B21, available on Zenodo\footnote{https://zenodo.org/records/12802110}).

\begin{figure}[h!]
    \centering
    \includegraphics[width=\columnwidth]{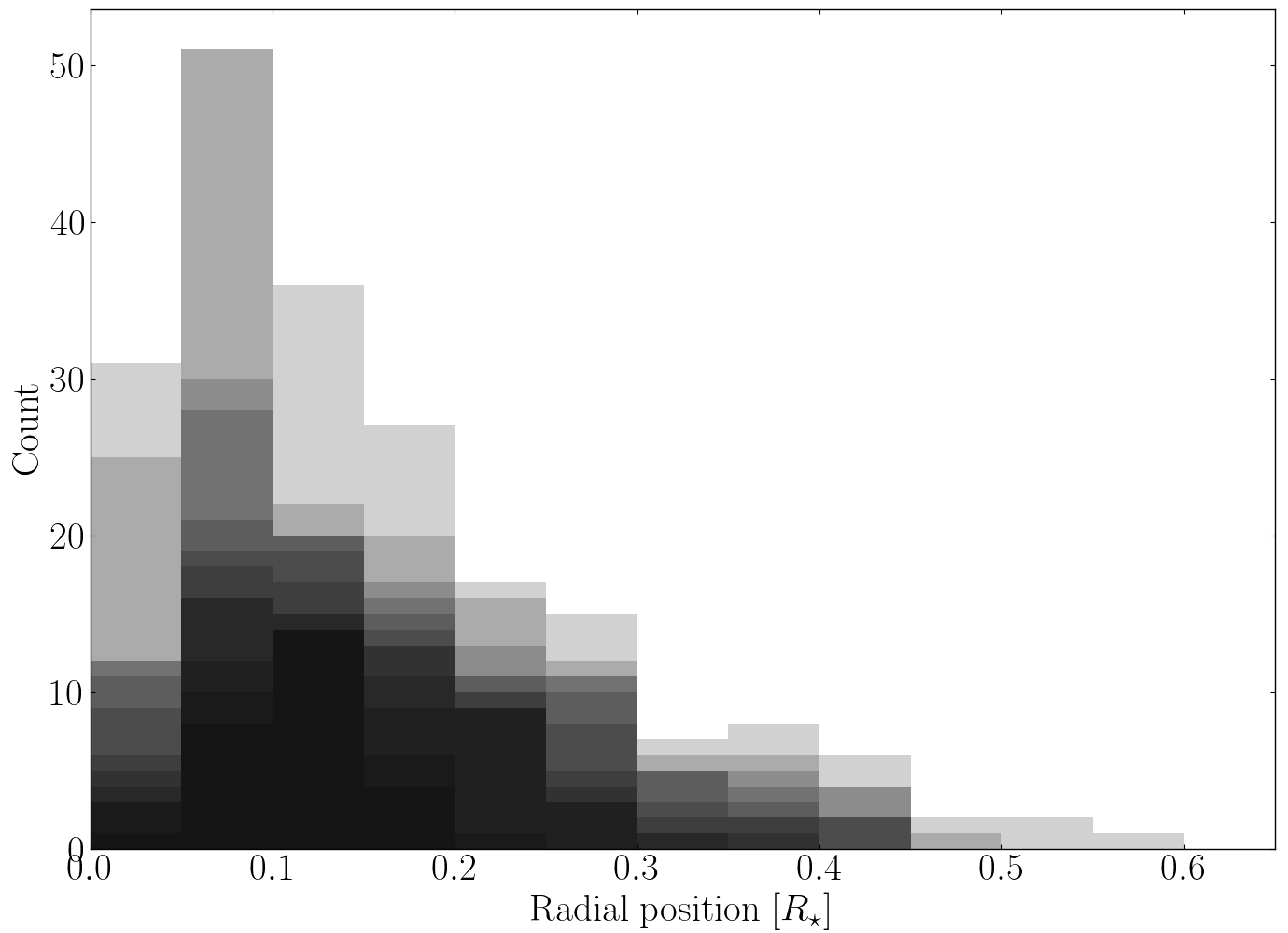}
    \caption{Histograms of the radial positions of the photocentre for all the intensity maps we computed and used in this work. The darker the shadow, the more often the photocentre is situated in the associated bin. Overall, we notice the photocentre is situated between $5\%$ and $15\%$ of the stellar radius.}
    \label{fig_hist_photo}
\end{figure}

We also computed the histogram of the radial position of the photocentre for every snapshot available of every simulation (Fig. \ref{fig_hist_photo}). The radial position is defined as $P=(P_x^2+P_y^2)^{1/2}$ in \% of $R_\star$. We notice that the photocentre is mainly situated between $0.05$ and $0.15$ of the stellar radius.

\subsection{Correlation between the photocentre displacement and the stellar parameters}\label{Sec:correlation}

After the correlations between $\mathrm{P_{puls}}$ and the stellar parameters ($\log(\mathrm{g})$, $\mathrm{T_{eff}}$ and $\mathrm{R_\star}$), we studied correlations between $\mathrm{P_{puls}}$ and the photocentre displacement, $\mathrm{\sigma_P}$, displayed in Fig. \ref{fig_relations}. The pulsation period increases when the surface gravity decreases (Fig. \ref{fig_logg_ppsul}); in other words, when the stellar radius increases. The photocentre gets displaced across larger distances \citep{chiavassa_heading_2018}. We notice a correlation for each sub-group that we approximated with a power law, whose parameters were determined with a non-linear least-squares method (see Eq. (\ref{formule_sP_logPpuls_1M}) and (\ref{formule_sP_logPpuls_15M})):

\begin{figure}[h!]
    \centering
    \includegraphics[width=\columnwidth]{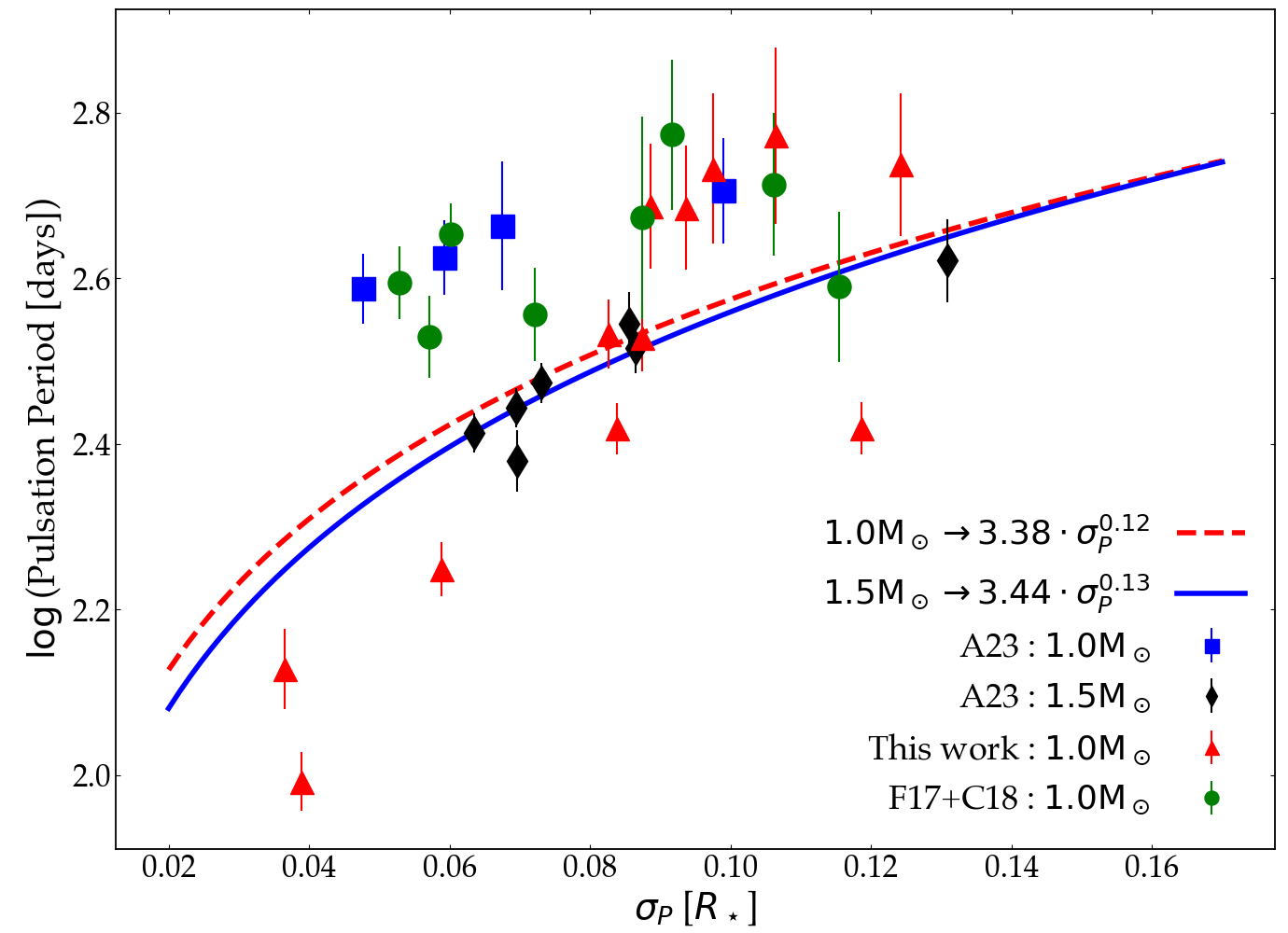}
    \caption{Pulsation period in logarithmic scale versus the photocentre displacement for the $31$ models. We group the models whether their mass is equal to $\mathrm{1.0\, M_\odot}$ or $\mathrm{1.5\, M_\odot}$. We approximate our data by a power law, one for each group (see Eq. (\ref{formule_sP_logPpuls_1M}), (\ref{formule_sP_logPpuls_15M})).}
    \label{fig_relations}
\end{figure}

\begin{equation}
    \log(\mathrm{P_{puls}})\, = \,3.38 \cdot x^{0.12}\mathrm{,\ \ with\ }x\mathrm{\, =\, \sigma_P\ or\ }x\mathrm{\, = \,\sigma_\varpi}
    \label{formule_sP_logPpuls_1M}
\end{equation}

\begin{equation}
\log(\mathrm{P_{puls}})\, =\, 3.44 \cdot x^{0.13}\mathrm{,\ \ with\ }x\mathrm{\, =\, \sigma_P\ or\ }x\mathrm{\, = \,\sigma_\varpi}
    \label{formule_sP_logPpuls_15M}
.\end{equation}

The resulting reduced $\bar{\chi}^2$ are $\bar{\chi}_{1.0}^{2} \sim 161$ and $\bar{\chi}_{1.5}^{2} \sim 2$. As before, we note a stark difference because there are fewer $\mathrm{1.5\,  M_\odot}$ simulations and they are less scattered.

\section{Comparison with observations from the \textit{Gaia} mission}

From the RHD simulations, we found analytical laws between the pulsation period and stellar parameters and between the pulsation period and the photocentre displacement, which were used to estimate the uncertainties of the results. We used these laws with the parallax uncertainty, $\mathrm{\sigma_\varpi}$, measured by \textit{Gaia} to derive the stellar parameters of observed stars. 

\subsection{Selection of the sample}\label{Sec:threeone}

To compare the analytical laws with observational data, the parameters of the observed stars need to match the parameters of the simulations. \cite{uttenthaler_interplay_2019} investigated the interplay between mass-loss and third dredge-up (3DUP) of a sample of variable stars in the solar neighbourhood, which we further constrained to select suitable stars for our analysis by following these conditions: (i) Mira stars with an assumed solar metallicity, (ii) a luminosity, $\mathrm{L_\star}$, lower than $\mathrm{10000 L_\odot}$, and $\mathrm{\sigma_{L_\star} / L_\star} < 50 \% $, $\mathrm{\sigma_{L_\star}}$ being the uncertainty on the luminosity, and (iii) the GDR3 parallax uncertainty, $\mathrm{\sigma_\varpi}$, lower than $0.14$ mas. We also selected stars whose (iv) renormalised unit weight error (RUWE) is lower than $1.4$ \citep{andriantsaralaza_distance_2022}. This operation resulted in a sample of $53$ Mira stars (Table \ref{Table_Miras}).

The pulsation periods were taken mainly from \cite{templeton_secular_2005} where available, or were also collected from VizieR. Preference was given to sources with available light curves that allowed for a critical evaluation of the period, such as the All Sky Automated Survey \citep{pojmanski_all_1998}. Since some Miras have pulsation periods that change significantly in time \citep{wood_helium-shell_1981,templeton_secular_2005}, we also analysed visual photometry from the AAVSO\footnote{https://www.aavso.org/} database and determined present-day periods with the program Period04\footnote{http://period04.net/} \citep{lenz_period04_2005}.
The analysis of \cite{merchan-benitez_meandering_2023} provided a period variability of the order of $2.4 \%$ of the respective pulsation period for Miras in the solar neighbourhood.

The luminosities were determined from a numerical integration under the photometric spectral energy distribution between the B-band at the short end and the IRAS $60\, \mu m$ band or, if available, the Akari $90\, \mu m$ band. A linear extrapolation to $\lambda=0$ and $\nu=0$ was taken into account. The photometry was corrected for interstellar extinction using the map of \cite{gontcharov_3d_2017}. We adopted the GEDR3 parallaxes and applied the average zero-point offset of quasars found by \cite{lindegren_gaia_2021}: $-21\, mas$.
Two main sources of uncertainty on the luminosity are the parallax uncertainty and the intrinsic variability of the stars. The uncertainty on interstellar extinction and on the parallax zero-point were neglected. 

The RUWE\footnote{https://gea.esac.esa.int/archive/documentation/GDR2/, Part V, Chapter 14.1.2} is expected to be around $1.0$ for sources where the single-star model provides a good fit to the astrometric observations. Following \cite{andriantsaralaza_distance_2022}, we rejected stars whose RUWE is above $1.4$ as their astrometric solution is expected to be poorly reliable and may indicate unresolved binaries. 

\begin{figure}
    \centering
    \includegraphics[width=\columnwidth]{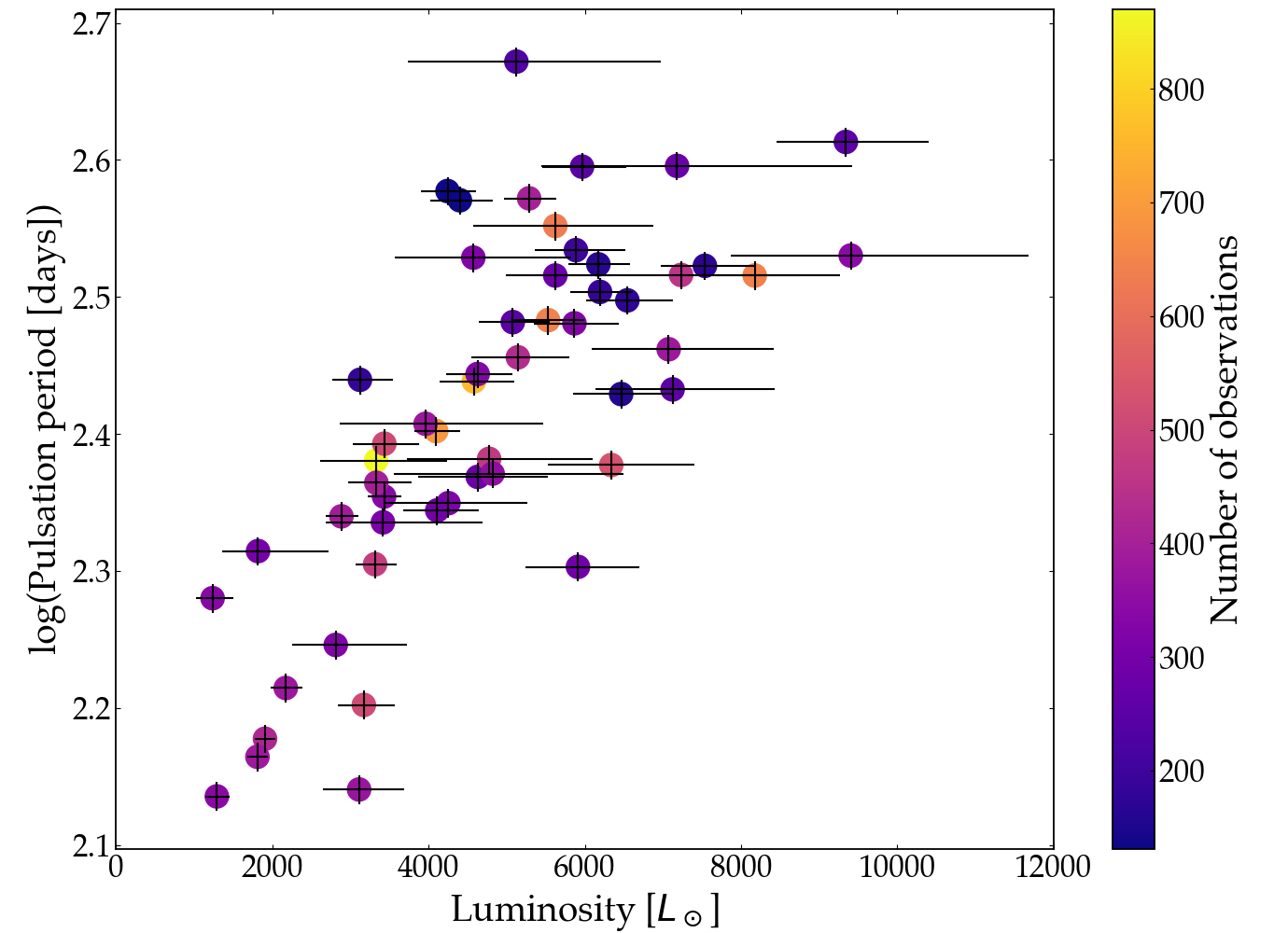}
    \includegraphics[width=\columnwidth]{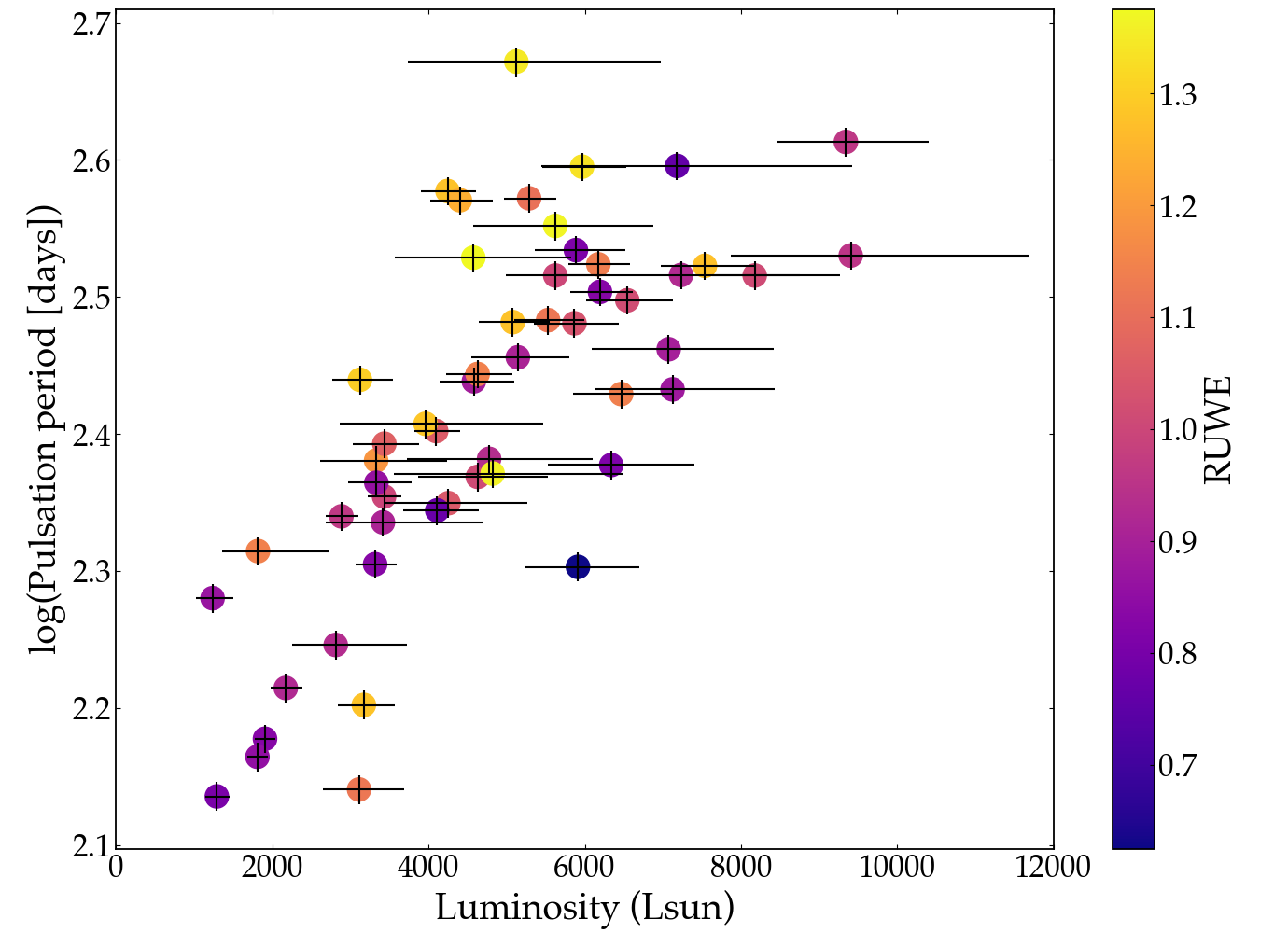}
    \caption{Pulsation period in logarithmic scale [days] of the sample versus their luminosity, [$L_\odot$]. \textbf{Top panel:} Number of observations in colour scale. \textbf{Bottom panel:} RUWE in colour scale.}
    \label{fig_RUWE_dist}
\end{figure}

Figure \ref{fig_RUWE_dist} displays the location of stars in a pulsation period-luminosity diagram with the colour scale indicating the number of good along-scan observations — that is, \textit{astrometric\_n\_good\_obs\_al} data from GDR3 (top panel) — or indicating the RUWE (bottom panel).
We investigated whether the number of times a star is observed, $\mathrm{N_{obs}}$, or the number of observed periods, $\mathrm{N_{per}}$, is correlated with RUWE. We do not see improvements of the astrometric solution when more observations are used to compute it. 

Information about the third dredge-up activity of the stars is available from spectroscopic observations of the absorption lines of technetium \citep[Tc, see][]{uttenthaler_interplay_2019}. Tc-rich stars have undergone a third dredge-up event and are thus more evolved and/or more massive than Tc-poor stars. Also, since $^{12}C$ is dredged up along with Tc, the C/O ratio could be somewhat enhanced in the Tc-rich compared to the Tc-poor stars. However, as our subsequent analysis showed, we do not find significant differences between Tc-poor and Tc-rich stars with respect to their astrometric characteristics; hence, this does not impact our results.

\subsection{Origins of the parallax uncertainty}

The uncertainty in \textit{Gaia} parallax measurements has multiple origins: (i) instrumental \citep{lindegren_gaia_2021}, (ii) distance \citep{lindegren_gaia_2021}, and (iii) convection-related \citep{chiavassa_pasquato_2011,chiavassa_heading_2018,chiavassa_probing_2022}. This makes the error budget difficult to estimate. In particular, only with time-dependent parallaxes with \textit{Gaia} Data Release 4 will the convection-related part be definitively characterised \citep{chiavassa_atmospheric_2019}. 
Concerning the distance and instrumental parts, \cite{lindegren_gaia_2021} investigated the bias of the parallax versus magnitude, colour, and position and developed an analytic method to correct the parallax of these biases. For comparison, the parallax and the corrected parallax are displayed in Table \ref{Table_Miras}, columns 8 and 9, respectively.
In this work, we assume that convective-related variability is the main contributor to the parallax uncertainty budget, which is already hinted at in observations; thus, $\mathrm{\sigma_P}$ is equivalent to $\mathrm{\sigma_\varpi}$. 

Indeed, optical interferometric observations of an AGB star showed the presence of large convective cells that affected the photocentre position. It has been shown for the same star that the convection-related variability accounts for a substantial part of the \textit{Gaia} Data Release 2 parallax error \citep[see in particular Fig. 2, ][]{chiavassa_optical_2020}. 

\subsection{Retrieval of the surface gravity based on the analytical laws}\label{Sec:threethree}

In Section~\ref{Sec:correlation}, we established an analytical law, Eq. (\ref{formule_sP_logPpuls_1M}),  between $\mathrm{\sigma_P}$ and $\mathrm{P_{puls}}$ with the $\mathrm{1.0\, M_\odot}$ simulations. We used it to calculate the pulsation period, $\mathrm{P_{1.0}}$, of the stars from observed $\mathrm{\sigma_\varpi}$. We defined $\mathrm{\Delta P_{1.0}}$, the relative difference between the observed pulsation period, $\mathrm{P_{obs}}$, and our results as $\mathrm{\Delta P_{1.0}}$\, = \,$\frac{\mathrm{P_{obs}}-\mathrm{P_{1.0}}}{\mathrm{P_{obs}}}$. This intermediate step gives an estimation of the error when computing the pulsation period and comparing it with observations. This error can then be used as a guideline to estimate the uncertainties of the final results; in other words, of the effective temperature, the surface gravity, and the radius of the stars in our sample.

\begin{figure}[h!]
    \centering
    \includegraphics[width=\columnwidth]{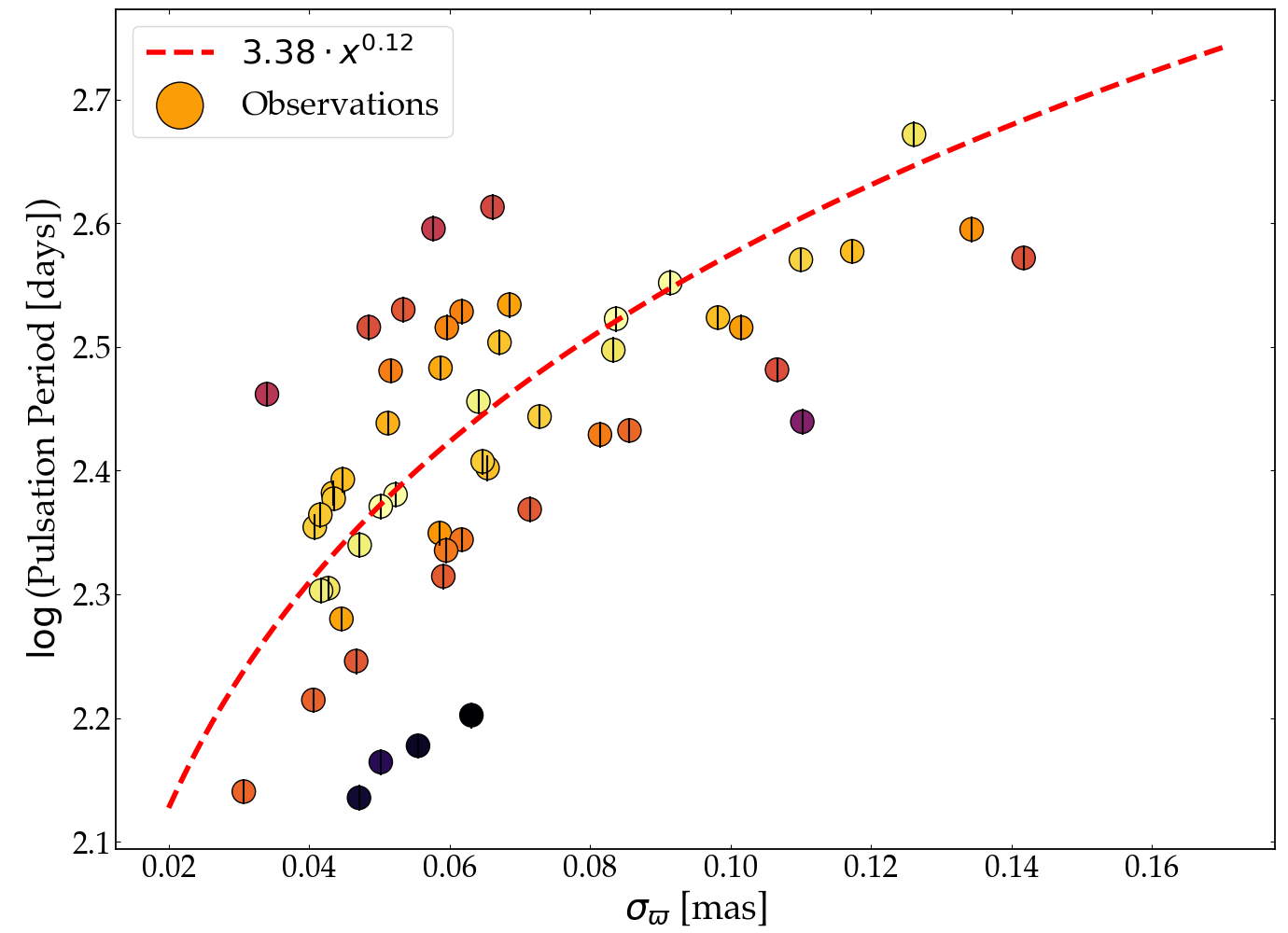}
    \includegraphics[width=\columnwidth]{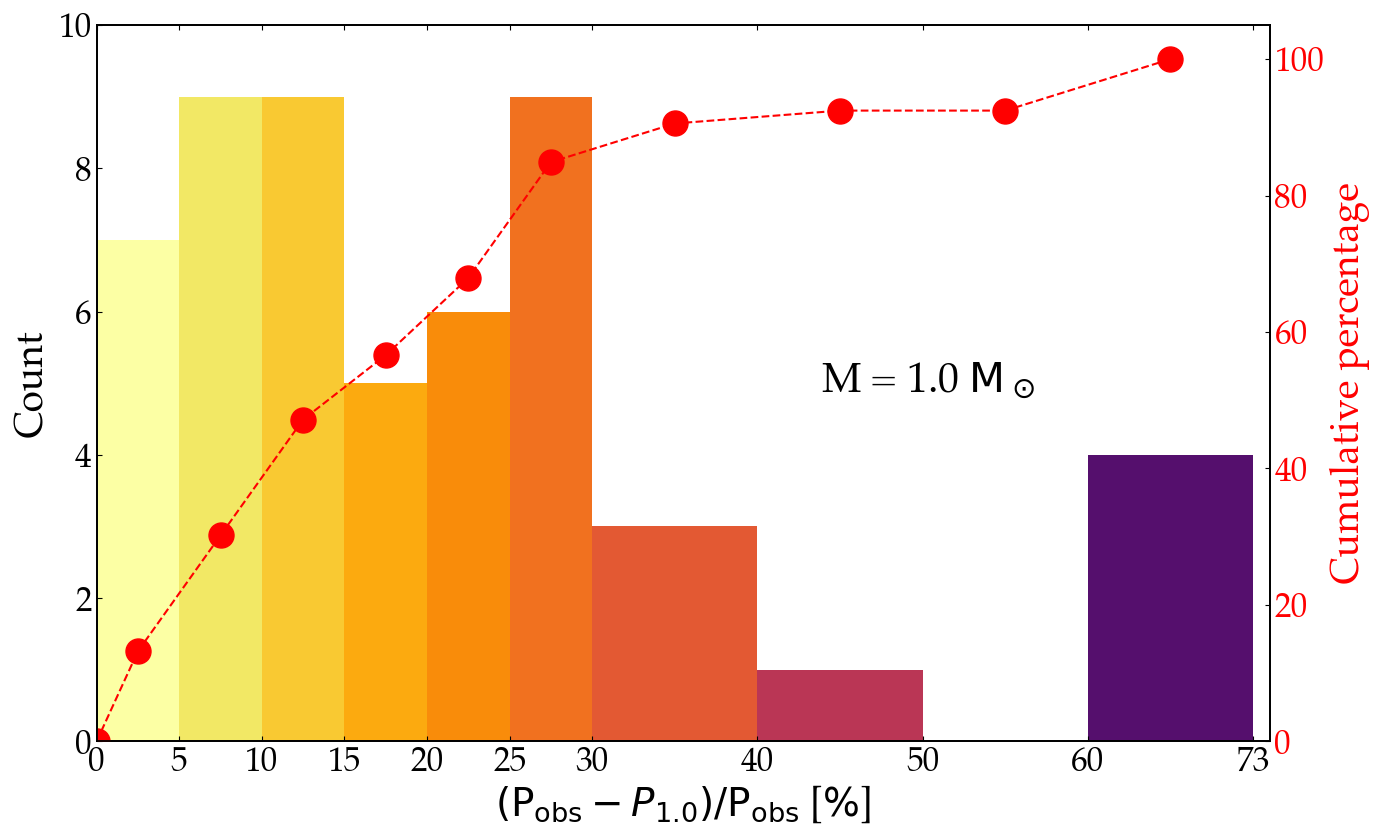}
    \caption{Comparison of the pulsation period of the sample between observations and estimations from the simulations where $M_\star\, = \,1.0\, M_\odot$. \textbf{Top panel:} Pulsation period, $\mathrm{P_{obs}}$, in logarithmic scale versus the parallax uncertainty of the observed sample, the dashed red curve being the Eq. (\ref{formule_sP_logPpuls_1M}) inferred from the analysis of the $\mathrm{1.0\, M_\odot}$ simulations. The colours represent the relative difference between the pulsation period calculated from Eq. (\ref{formule_sP_logPpuls_1M}) and the observations. \textbf{Bottom panel}: Histogram of these relative differences. The red line accounts for the cumulative number of stars in the respective and preceding bins (see right-hand Y-axis). The limits of the last bin are $60 \%$ and $73 \%$, with only one star above $70 \%$}. 
    \label{fig_eplx_logppuls_obs_10M}
\end{figure}

The top panel of Figure \ref{fig_eplx_logppuls_obs_10M} displays  $\mathrm{P_{obs}}$ versus $\mathrm{\sigma_\varpi}$ as dots and $\mathrm{P_{1.0}}$ versus $\mathrm{\sigma_\varpi}$ as the dashed red curve. 
The bottom panel displays the histogram of the relative difference, $\mathrm{\Delta P_{1.0}}$, and the cumulative percentage of the observed sample. The same colour scale is used in both panels: for example, light yellow represents a relative difference between $\mathrm{P_{1.0}}$ and $\mathrm{P_{obs}}$ of less than $5 \%$, while purple represents a relative difference greater than $60 \%$ (column 3 in Table \ref{Table_Miras_RES}). 

$\mathrm{\Delta P_{1.0}}$ ranges from $0.4 \%$ to $72 \%$, with a median of $16 \%$ (i.e. from $1$ to $68$ days' difference). For $85 \%$ of the sample stars, $\mathrm{\Delta P_{1.0}}$ is $\le 30 \%$, and for $57 \%$, it is $\le 20 \%$, suggesting we statistically have a good agreement between our model results and the observations.

\begin{figure}[h!]
    \centering
    \includegraphics[width=\columnwidth]{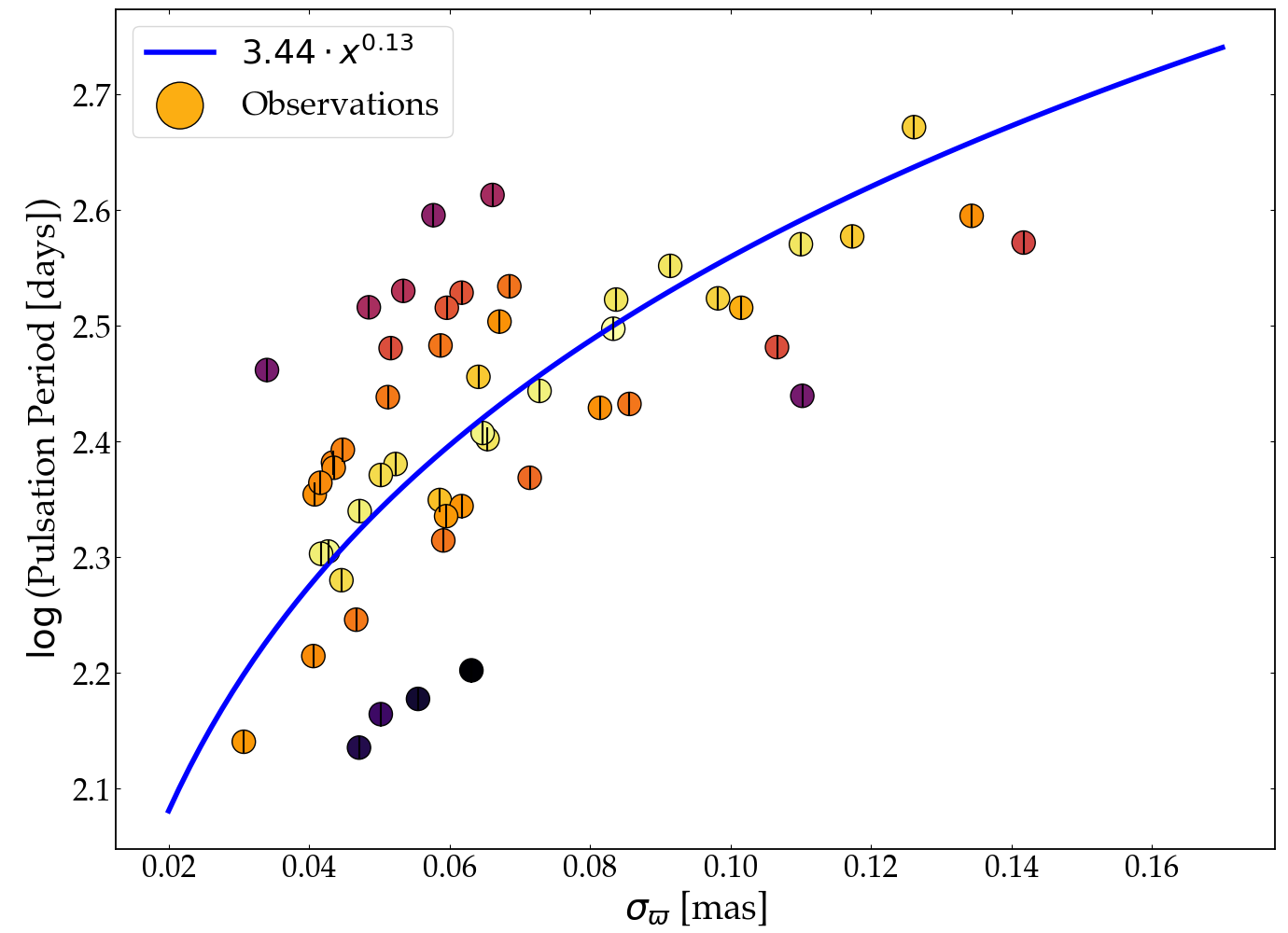}
    \includegraphics[width=\columnwidth]{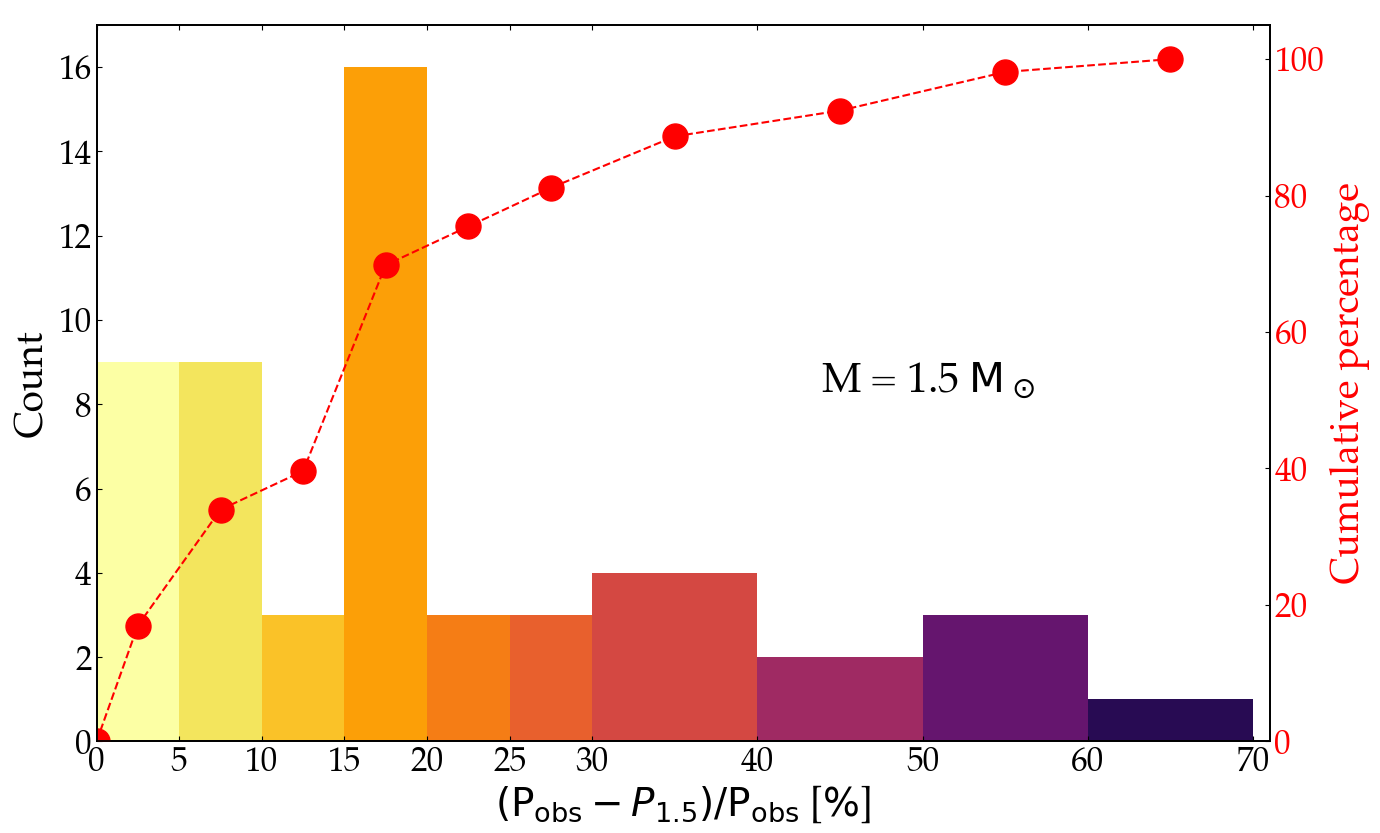}
    \caption{Comparison of the pulsation period of the sample between observations and estimations from the simulations where $M_\star\, = \,1.5\, M_\odot$. \textbf{Top panel:} Pulsation period, $\mathrm{P_{obs}}$, in logarithmic scale versus the parallax uncertainty of the observed sample, the blue curve being the Eq. (\ref{formule_sP_logPpuls_15M}) inferred from the analysis of the $\mathrm{1.5\, M_\odot}$ simulations. The colours represent the relative difference between the pulsation period calculated from Eq. (\ref{formule_sP_logPpuls_15M} and with the observations. \textbf{Bottom panel}: Histogram of these relative differences. The red line accounts for the cumulative number of stars in the respective and preceding bins (see right-hand Y-axis).}
    \label{fig_eplx_logppuls_obs_15M}
\end{figure}

\begin{figure*}
\centering
\includegraphics[width=1.\textwidth]{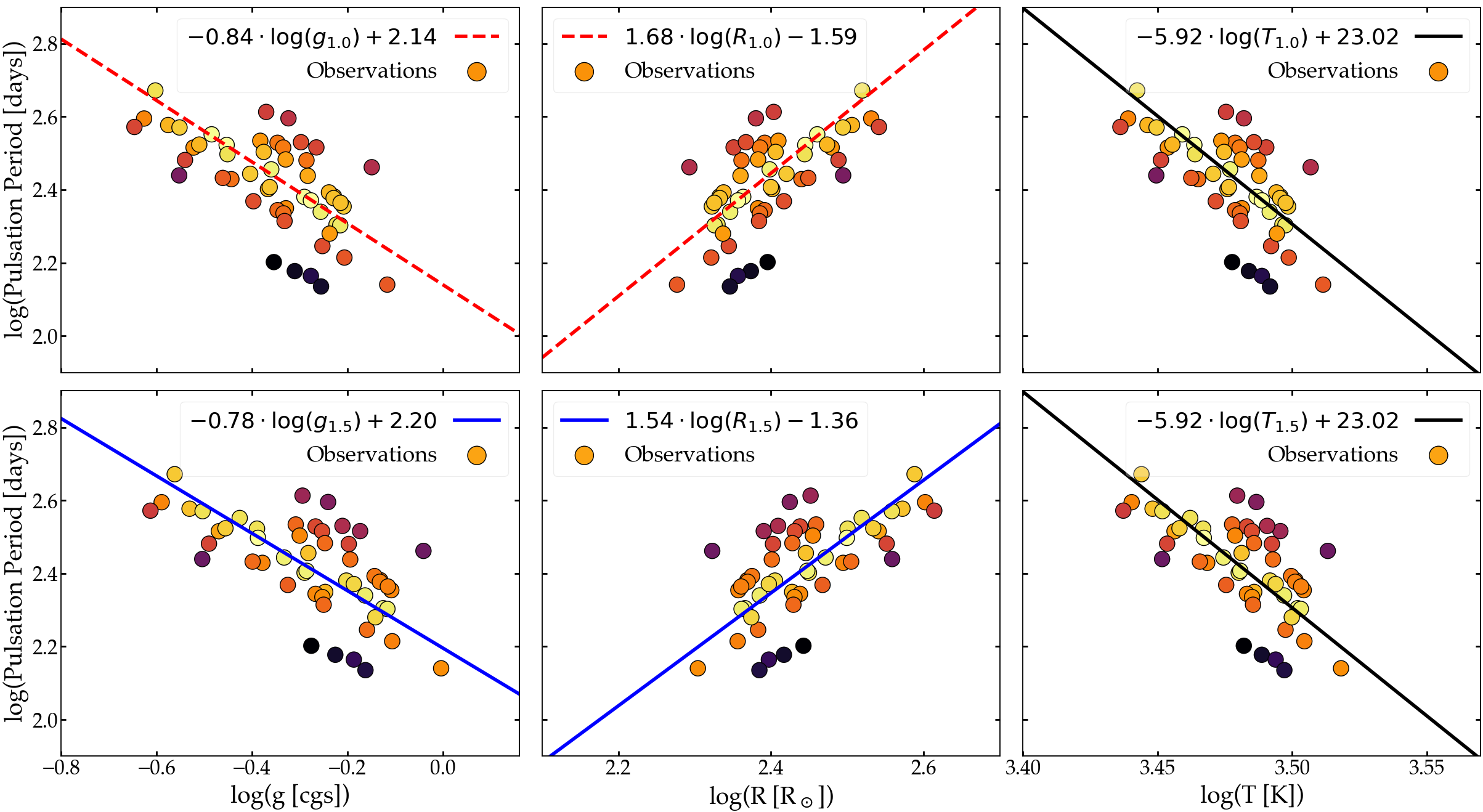}
\caption{Pulsation period from the observations versus stellar parameter values obtained from the simulations. \textbf{Top row: }Pulsation period versus $\log(\mathrm{g})$ (left column), $\log(\mathrm{R_\star})$ (central column), $\log(\mathrm{T_{eff}})$ (right column), each computed thanks to the analytical laws derived from the $\mathrm{1.0\, M_\odot}$ simulations. \textbf{Bottom row: }Same as in the top row for the $\mathrm{1.5\, M_\odot}$ simulations. The equation of each law is given in the legend and is reported in Fig. \ref{diagram}. The colour scale used is the same as in Figs. \ref{fig_eplx_logppuls_obs_10M} and \ref{fig_eplx_logppuls_obs_15M}.}
\label{fig_resultats} 
\end{figure*}

We then combined Eqs. (\ref{formule_sP_logPpuls_1M}) and (\ref{formule_logg_logPpuls10}) to derive the surface gravity ($\mathrm{\log(g_{1.0})}$) directly from $\mathrm{\sigma_\varpi}$ (column 6 in Table \ref{Table_Miras_RES}). The top left panel of Figure \ref{fig_resultats} displays $\mathrm{\log(P_{obs})}$ versus $\mathrm{\log(g_{1.0})}$ and Eq. (\ref{formule_logg_logPpuls10}) as the dashed red curve. We notice that the calculated $\mathrm{\log(g_{1.0})}$ values follow the same trend as $\mathrm{\log(g)}$ from the simulations and are the most accurate when closest to the line.

We performed the same analysis with the analytical laws from the $\mathrm{1.5\, M_\odot}$ simulations. The top panel of Figure \ref{fig_eplx_logppuls_obs_15M} displays the calculated $\mathrm{P_{1.5}}$ versus $\mathrm{\sigma_\varpi}$ and the bottom panel displays a histogram of the relative difference between $\mathrm{P_{1.5}}$ and $\mathrm{P_{obs}}$ defined as $\mathrm{\Delta P_{1.5}}$ (column 5 in Table \ref{Table_Miras_RES}). The same colour code is used in both panels. We combined Eqs. ($\ref{formule_sP_logPpuls_15M}$) and ($\ref{formule_logg_logPpuls15}$) to compute $\mathrm{\log(g_{1.5})}$. Fig. \ref{fig_resultats} displays $\mathrm{\log(P_{obs})}$ from the simulations versus $\mathrm{\log(g_{1.5})}$.

For the $\mathrm{1.5\, M_\odot}$ simulations, $\mathrm{\Delta P_{1.5}}$ ranges from $0.6 \%$ to $62 \%$, with a median of $16 \%$ (i.e. from $1$ to $76$ days' difference). For $81 \%$ of the sample stars, $\mathrm{\Delta P_{1.5}}$ is $\le 30 \%$, and for $70 \%$, it is $\le 20 \%$.

Qualitatively, we observe two different analytical laws depending on the mass of the models used to infer the laws. However, a larger grid of simulations, covering the mass range of AGB stars, would help to confirm this trend, tailor the analytical laws to specific stars, and predict more precise stellar parameters.

\subsection{Retrieval of the other stellar parameters}

We repeated the same procedure to retrieve the stellar radius, $\mathrm{R_{1.0}}$, and the effective temperature, $\mathrm{T_{1.0}}$ (columns 8 and 10 in Table \ref{Table_Miras_RES}). Combining Eqs. (\ref{formule_sP_logPpuls_1M}) and (\ref{formule_logR_logPpuls_10M}), we computed the radius, $\mathrm{R_{1.0}}$. The top central panel of Figure \ref{fig_resultats} displays $\mathrm{\log(P_{obs})}$ versus $\mathrm{\log(R_{1.0})}$. 
Combining the Eqs. (\ref{formule_sP_logPpuls_1M}) and (\ref{formule_logT_logPpuls}), we computed the effective temperature, $\mathrm{T_{1.0}}$ (Fig. \ref{fig_resultats}, top right panel). 
As in Section~\ref{Sec:threethree} for $\mathrm{\log(g_{1.0})}$, the calculated $\mathrm{\log(R_{1.0})}$ follow the Eq. \ref{formule_logR_logPpuls_10M} derived from the simulations, and $\mathrm{\log(R_{1.0})}$ follow Eq. \ref{formule_logT_logPpuls}.

We repeated the same procedure for the $\mathrm{1.5\, M_\odot}$ simulations to retrieve $\mathrm{R_{1.5}}$ and $\mathrm{T_{1.5}}$ (columns 9 and 11 in Table \ref{tab_simus}). The results are displayed in the bottom central and right panels of Fig. \ref{fig_resultats}.

\begin{figure*}[!h]
    \centering
    \includegraphics[width=0.8\textwidth]{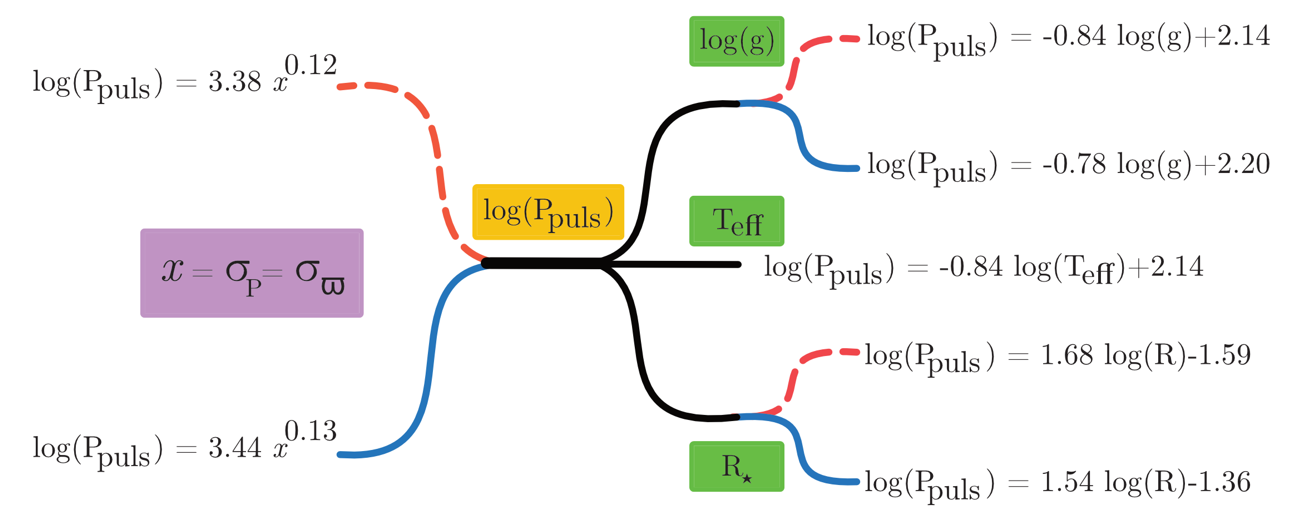}
    \caption{Summary of the analytical laws we established. The dashed red lines correspond to the analysis when using only the $\mathrm{1.0\, M_\odot}$ simulations, the blue lines the analysis when using only the $\mathrm{1.5\, M_\odot}$ simulations. For the correlation between the pulsation period and the effective temperature, all available simulations were used. In the left part, we have the laws between the photocentre displacement and the pulsation period from the simulations. We assume that the parallax uncertainty, $\mathrm{\sigma_\varpi}$, is equivalent to $\mathrm{\sigma_P}$ as $\mathrm{\sigma_P}$ is the main contributor of the $\mathrm{\sigma_\varpi}$ budget. In the right part, we have the laws between the pulsation period and the stellar parameters ($\mathrm{\log(g)}$, $\mathrm{T_{eff}}$, $\mathrm{R_\star}$). Combining these laws provides the parameters of the Mira stars based on the parallax uncertainty. The relative difference between the calculated and the observational pulsation periods can be used to estimate the uncertainties of the stellar parameters.}
    \label{diagram}
\end{figure*}

For comparison, R Peg has been observed with the interferometric instrument GRAVITY/VLTI in 2017, which provided a direct estimation of its radius: $\mathrm{R_{Ross}}\, =\, 351^{+38}_{-31}\, \mathrm{R_\odot}$ \citep{wittkowski_vlti-gravity_2018}. From our study, $\mathrm{R_{1.0,RPeg}}\, =\, 321^{+5}_{-15}\, \mathrm{R_\odot}$ and $\mathrm{R_{1.5,RPeg}}\, =\, 373^{+5}_{-14}\, \mathrm{R_\odot}$, which is in good agreement with the results of \cite{wittkowski_vlti-gravity_2018}. With future interferometric observations, we shall be able to further validate our results. 

\section{Summary and conclusions}

We computed intensity maps in the \textit{Gaia} band from the snapshots of $31$ RHD simulations of AGB stars computed with CO$^5$BOLD. The standard deviation of the photocentre displacement, $\mathrm{\sigma_P}$, due to the presence of large convective cells on the surface, ranges from about $4 \%$ to $13 \%$ of its corresponding stellar radius, which is coherent with previous studies and is non-negligible in photometric data analysis. It becomes the main contributor to the \textit{Gaia} parallax uncertainty $\mathrm{\sigma_\varpi}$ budget. The dynamics and winds of the AGB stars also affect the determination of stellar parameters and amplify their uncertainties. It becomes worth exploring the correlations between all these aspects to eventually retrieve such parameters from the parallax uncertainty.

We provided correlations between the photocentre displacement and the pulsation period as well as between the pulsation period and stellar parameters: the effective surface gravity, $\mathrm{\log(g)}$, the effective temperature, $\mathrm{T_{eff}}$, and the radius, $\mathrm{R_\star}$. We separated the simulations into two sub-groups based on whether their mass is equal to $1.0$ or $\mathrm{1.5\, M_\odot}$. Indeed, the laws we provided, and the final results, are sensitive to the mass. A grid of simulations covering a larger range of masses, with a meaningful number of simulations for each, would help confirm this observation and establish laws that are suitable for varied stars. This will be done in the future.

We then applied these laws to a sample of $53$ Mira stars matching the simulations' parameters. We first compared the pulsation period with the literature: we obtained a relative error of less than $30 \%$ for $85 \%$ of the stars in the sample for the first case and $81 \%$ for the second, which indicates reasonable results from a statistical point of view. This error can then be used as a guideline to estimate the uncertainties of the final results. We then computed $\mathrm{\log(g)}$, $\mathrm{T_{eff}}$ and $\mathrm{R_\star}$ by combining the analytical laws (Table \ref{Table_Miras_RES}). 

While mass loss from red giant branch stars should be mainly independent of metallicity, it has been suggested that this is less true for AGB stars \citep{mcdonald_mass-loss_2015}. Photocentre displacement and stellar parameters may be dependent on metallicity and this question needs to be further studied.

We argue that the method used for Mira stars presented in this article, based on RHD simulations, can be generalised to any AGB stars whose luminosity is in the $2000$--$10000\, \mathrm{L_\odot}$ range and that have a \textit{Gaia} parallax uncertainty below $0.14$ mas. Figure \ref{diagram} sums up the analytical laws found in this work that can be used to calculate the stellar parameters.

Overall, we have demonstrated the feasibility of retrieving stellar parameters for AGB stars using their uncertainty on the parallax, thanks to the employment of state-of-the-art 3D RHD simulations of stellar convection. The future \textit{Gaia} Data Release 4 will provide time-dependent parallax measurements, allowing one to quantitatively determine the photocentre-related impact on the parallax error budget and to directly compare the convection cycle, refining our understanding of AGB dynamics. 

\begin{acknowledgements}
This work is funded by the French National Research Agency (ANR) project PEPPER (ANR-20-CE31-0002). BF acknowledges funding from the European Research Council (ERC) under the European Union’s Horizon 2020 research and innovation programme Grant agreement No. 883867, project EXWINGS. The computations were enabled by resources provided by the Swedish National Infrastructure for Computing (SNIC). This work was granted access to the HPC resources of Observatoire de la Côte d’Azur - Mésocentre SIGAMM.
\end{acknowledgements}

\bibliographystyle{aa} 
\bibliography{bib.bib} 

\begin{thebibliography}{34}
\expandafter\ifx\csname natexlab\endcsname\relax\def\natexlab#1{#1}\fi

\bibitem[{Ahmad {et~al.}(2023)Ahmad, Freytag, \& Höfner}]{ahmad_properties_2023}
Ahmad, A., Freytag, B., \& Höfner, S. 2023, \aap, 669, A49

\bibitem[{Andriantsaralaza {et~al.}(2022)Andriantsaralaza, Ramstedt, Vlemmings, \& De~Beck}]{andriantsaralaza_distance_2022}
Andriantsaralaza, M., Ramstedt, S., Vlemmings, W. H.~T., \& De~Beck, E. 2022, \aap, 667, A74

\bibitem[{Asplund {et~al.}(2009)Asplund, Grevesse, Sauval, \& Scott}]{asplund_chemical_2009}
Asplund, M., Grevesse, N., Sauval, A.~J., \& Scott, P. 2009, \araa, 47, 481

\bibitem[{Chen {et~al.}(2021)Chen, Xie, Zhou, Dong, Liu, Wang, Xiang, Huang, Luo, \& Zheng}]{chen_planets_2021}
Chen, D.-C., Xie, J.-W., Zhou, J.-L., {et~al.} 2021, \apj, 909, 115

\bibitem[{Chiavassa {et~al.}(2018)Chiavassa, Freytag, \& Schultheis}]{chiavassa_heading_2018}
Chiavassa, A., Freytag, B., \& Schultheis, M. 2018, \aap, 617, L1

\bibitem[{Chiavassa {et~al.}(2019)Chiavassa, Freytag, \& Schultheis}]{chiavassa_atmospheric_2019}
Chiavassa, A., Freytag, B., \& Schultheis, M. 2019, Proceedings of the Annual meeting of the French Society of Astronomy and Astrophysics

\bibitem[{Chiavassa {et~al.}(2020)Chiavassa, Kravchenko, Millour, Schaefer, Schultheis, Freytag, Creevey, Hocdé, Morand, Ligi, Kraus, Monnier, Mourard, Nardetto, Anugu, Le~Bouquin, Davies, Ennis, Gardner, Labdon, Lanthermann, Setterholm, \& ten Brummelaar}]{chiavassa_optical_2020}
Chiavassa, A., Kravchenko, K., Millour, F., {et~al.} 2020, \aap, 640, A23

\bibitem[{Chiavassa {et~al.}(2022)Chiavassa, Kudritzki, Davies, Freytag, \& de~Mink}]{chiavassa_probing_2022}
Chiavassa, A., Kudritzki, R., Davies, B., Freytag, B., \& de~Mink, S.~E. 2022, \aap, 661, L1

\bibitem[{Chiavassa {et~al.}(2011)Chiavassa, Pasquato, Jorissen, Sacuto, Babusiaux, Freytag, Ludwig, Cruzalèbes, Rabbia, Spang, \& Chesneau}]{chiavassa_pasquato_2011}
Chiavassa, A., Pasquato, E., Jorissen, A., {et~al.} 2011, \aap, 528, A120

\bibitem[{Chiavassa {et~al.}(2009)Chiavassa, Plez, Josselin, \& Freytag}]{chiavassa_radiative_2009}
Chiavassa, A., Plez, B., Josselin, E., \& Freytag, B. 2009, \aap, 506, 1351

\bibitem[{De~Beck {et~al.}(2010)De~Beck, Decin, de~Koter, Justtanont, Verhoelst, Kemper, \& Menten}]{de_beck_probing_2010}
De~Beck, E., Decin, L., de~Koter, A., {et~al.} 2010, \aap, 523, A18

\bibitem[{Decin(2021)}]{decin_evolution_2021}
Decin, L. 2021, \araa, 59, 337

\bibitem[{Freytag(2013)}]{freytag_advances_2013}
Freytag, B. 2013, Mem. Soc. Astron. Ital. Suppl., 24, 26

\bibitem[{Freytag(2017)}]{freytag_boundary_2017}
Freytag, B. 2017, Mem. Soc. Astron. Ital., 88, 12

\bibitem[{Freytag {et~al.}(1997)Freytag, Holweger, Steffen, \& Ludwig}]{freytag_scale_1997}
Freytag, B., Holweger, H., Steffen, M., \& Ludwig, H.~G. 1997, Proceedings of the ESO Workshop, 316

\bibitem[{Freytag \& Höfner(2023)}]{freytag_global_2023}
Freytag, B. \& Höfner, S. 2023, \aap, 669, A155

\bibitem[{Freytag {et~al.}(2017)Freytag, Liljegren, \& Höfner}]{freytag_global_2017}
Freytag, B., Liljegren, S., \& Höfner, S. 2017, \aap, 600, A137

\bibitem[{Freytag {et~al.}(2012)Freytag, Steffen, Ludwig, Wedemeyer-Böhm, Schaffenberger, \& Steiner}]{freytag_simulations_2012}
Freytag, B., Steffen, M., Ludwig, H.~G., {et~al.} 2012, J. Comput. Phys., 231, 919

\bibitem[{{Gaia Collaboration} {et~al.}(2016){Gaia Collaboration}, Prusti, de~Bruijne, Brown, Vallenari, Babusiaux, Bailer-Jones, Bastian, Biermann, Evans, Eyer, Jansen, Jordi, Klioner, Lammers, Lindegren, Luri, Mignard, Milligan, Panem, Poinsignon, Pourbaix, Randich, Sarri, Sartoretti, Siddiqui, Soubiran, Valette, van Leeuwen, Walton, Aerts, Arenou, Cropper, Drimmel, Høg, Katz, Lattanzi, O'Mullane, Grebel, Holland, Huc, Passot, Bramante, Cacciari, Castañeda, Chaoul, Cheek, De~Angeli, Fabricius, Guerra, Hernández, Jean-Antoine-Piccolo, Masana, Messineo, Mowlavi, Nienartowicz, Ordóñez-Blanco, Panuzzo, Portell, Richards, Riello, Seabroke, Tanga, Thévenin, Torra, Els, Gracia-Abril, Comoretto, Garcia-Reinaldos, Lock, Mercier, Altmann, Andrae, Astraatmadja, Bellas-Velidis, Benson, Berthier, Blomme, Busso, Carry, Cellino, Clementini, Cowell, Creevey, Cuypers, Davidson, De~Ridder, de~Torres, Delchambre, Dell'Oro, Ducourant, Frémat, García-Torres, Gosset, Halbwachs, Hambly, Harrison, Hauser, Hestroffer,
  Hodgkin, Huckle, Hutton, Jasniewicz, Jordan, Kontizas, Korn, Lanzafame, Manteiga, Moitinho, Muinonen, Osinde, Pancino, Pauwels, Petit, Recio-Blanco, Robin, Sarro, Siopis, Smith, Smith, Sozzetti, Thuillot, van Reeven, Viala, Abbas, Abreu~Aramburu, Accart, Aguado, Allan, Allasia, Altavilla, Álvarez, Alves, Anderson, Andrei, Anglada~Varela, Antiche, Antoja, Antón, Arcay, Atzei, Ayache, Bach, Baker, Balaguer-Núñez, Barache, Barata, Barbier, Barblan, Baroni, Barrado~y Navascués, Barros, Barstow, Becciani, Bellazzini, Bellei, Bello~García, Belokurov, Bendjoya, Berihuete, Bianchi, Bienaymé, Billebaud, Blagorodnova, Blanco-Cuaresma, Boch, Bombrun, Borrachero, Bouquillon, Bourda, Bouy, Bragaglia, Breddels, Brouillet, Brüsemeister, Bucciarelli, Budnik, Burgess, Burgon, Burlacu, Busonero, Buzzi, Caffau, Cambras, Campbell, Cancelliere, Cantat-Gaudin, Carlucci, Carrasco, Castellani, Charlot, Charnas, Charvet, Chassat, Chiavassa, Clotet, Cocozza, Collins, Collins, Costigan, Crifo, Cross, Crosta, Crowley, Dafonte,
  Damerdji, Dapergolas, David, David, De~Cat, de~Felice, de~Laverny, De~Luise, De~March, de~Martino, de~Souza, Debosscher, del Pozo, Delbo, Delgado, Delgado, di~Marco, Di~Matteo, Diakite, Distefano, Dolding, Dos~Anjos, Drazinos, Durán, Dzigan, Ecale, Edvardsson, Enke, Erdmann, Escolar, Espina, Evans, Eynard~Bontemps, Fabre, Fabrizio, Faigler, Falcão, Farràs~Casas, Faye, Federici, Fedorets, Fernández-Hernández, Fernique, Fienga, Figueras, Filippi, Findeisen, Fonti, Fouesneau, Fraile, Fraser, Fuchs, Furnell, Gai, Galleti, Galluccio, Garabato, García-Sedano, Garé, Garofalo, Garralda, Gavras, Gerssen, Geyer, Gilmore, Girona, Giuffrida, Gomes, González-Marcos, González-Núñez, González-Vidal, Granvik, Guerrier, Guillout, Guiraud, Gúrpide, Gutiérrez-Sánchez, Guy, Haigron, Hatzidimitriou, Haywood, Heiter, Helmi, Hobbs, Hofmann, Holl, Holland, Hunt, Hypki, Icardi, Irwin, Jevardat~de Fombelle, Jofré, Jonker, Jorissen, Julbe, Karampelas, Kochoska, Kohley, Kolenberg, Kontizas, Koposov, Kordopatis,
  Koubsky, Kowalczyk, Krone-Martins, Kudryashova, Kull, Bachchan, Lacoste-Seris, Lanza, Lavigne, Le~Poncin-Lafitte, Lebreton, Lebzelter, Leccia, Leclerc, Lecoeur-Taibi, Lemaitre, Lenhardt, Leroux, Liao, Licata, Lindstrøm, Lister, Livanou, Lobel, Löffler, López, Lopez-Lozano, Lorenz, Loureiro, MacDonald, Magalhães~Fernandes, Managau, Mann, Mantelet, Marchal, Marchant, Marconi, Marie, Marinoni, Marrese, Marschalkó, Marshall, Martín-Fleitas, Martino, Mary, Matijevič, Mazeh, McMillan, Messina, Mestre, Michalik, Millar, Miranda, Molina, Molinaro, Molinaro, Molnár, Moniez, Montegriffo, Monteiro, Mor, Mora, Morbidelli, Morel, Morgenthaler, Morley, Morris, Mulone, Muraveva, Musella, Narbonne, Nelemans, Nicastro, Noval, Ordénovic, Ordieres-Meré, Osborne, Pagani, Pagano, Pailler, Palacin, Palaversa, Parsons, Paulsen, Pecoraro, Pedrosa, Pentikäinen, Pereira, Pichon, Piersimoni, Pineau, Plachy, Plum, Poujoulet, Prša, Pulone, Ragaini, Rago, Rambaux, Ramos-Lerate, Ranalli, Rauw, Read, Regibo, Renk, Reylé,
  Ribeiro, Rimoldini, Ripepi, Riva, Rixon, Roelens, Romero-Gómez, Rowell, Royer, Rudolph, Ruiz-Dern, Sadowski, Sagristà~Sellés, Sahlmann, Salgado, Salguero, Sarasso, Savietto, Schnorhk, Schultheis, Sciacca, Segol, Segovia, Segransan, Serpell, Shih, Smareglia, Smart, Smith, Solano, Solitro, Sordo, Soria~Nieto, Souchay, Spagna, Spoto, Stampa, Steele, Steidelmüller, Stephenson, Stoev, Suess, Süveges, Surdej, Szabados, Szegedi-Elek, Tapiador, Taris, Tauran, Taylor, Teixeira, Terrett, Tingley, Trager, Turon, Ulla, Utrilla, Valentini, van Elteren, Van~Hemelryck, van Leeuwen, Varadi, Vecchiato, Veljanoski, Via, Vicente, Vogt, Voss, Votruba, Voutsinas, Walmsley, Weiler, Weingrill, Werner, Wevers, Whitehead, Wyrzykowski, Yoldas, Žerjal, Zucker, Zurbach, Zwitter, Alecu, Allen, Allende~Prieto, Amorim, Anglada-Escudé, Arsenijevic, Azaz, Balm, Beck, Bernstein, Bigot, Bijaoui, Blasco, Bonfigli, Bono, Boudreault, Bressan, Brown, Brunet, Bunclark, Buonanno, Butkevich, Carret, Carrion, Chemin, Chéreau, Corcione,
  Darmigny, de~Boer, de~Teodoro, de~Zeeuw, Delle~Luche, Domingues, Dubath, Fodor, Frézouls, Fries, Fustes, Fyfe, Gallardo, Gallegos, Gardiol, Gebran, Gomboc, Gómez, Grux, Gueguen, Heyrovsky, Hoar, Iannicola, Isasi~Parache, Janotto, Joliet, Jonckheere, Keil, Kim, Klagyivik, Klar, Knude, Kochukhov, Kolka, Kos, Kutka, Lainey, LeBouquin, Liu, Loreggia, Makarov, Marseille, Martayan, Martinez-Rubi, Massart, Meynadier, Mignot, Munari, Nguyen, Nordlander, Ocvirk, O'Flaherty, Olias~Sanz, Ortiz, Osorio, Oszkiewicz, Ouzounis, Palmer, Park, Pasquato, Peltzer, Peralta, Péturaud, Pieniluoma, Pigozzi, Poels, Prat, Prod'homme, Raison, Rebordao, Risquez, Rocca-Volmerange, Rosen, Ruiz-Fuertes, Russo, Sembay, Serraller~Vizcaino, Short, Siebert, Silva, Sinachopoulos, Slezak, Soffel, Sosnowska, Straižys, ter Linden, Terrell, Theil, Tiede, Troisi, Tsalmantza, Tur, Vaccari, Vachier, Valles, Van~Hamme, Veltz, Virtanen, Wallut, Wichmann, Wilkinson, Ziaeepour, \& Zschocke}]{gaia_collaboration_gaia_2016}
{Gaia Collaboration}, Prusti, T., de~Bruijne, J. H.~J., {et~al.} 2016, \aap, 595, A1

\bibitem[{{Gaia Collaboration} {et~al.}(2023){Gaia Collaboration}, Vallenari, Brown, Prusti, de~Bruijne, Arenou, Babusiaux, Biermann, Creevey, Ducourant, Evans, Eyer, Guerra, Hutton, Jordi, Klioner, Lammers, Lindegren, Luri, Mignard, Panem, Pourbaix, Randich, Sartoretti, Soubiran, Tanga, Walton, Bailer-Jones, Bastian, Drimmel, Jansen, Katz, Lattanzi, van Leeuwen, Bakker, Cacciari, Castañeda, De~Angeli, Fabricius, Fouesneau, Frémat, Galluccio, Guerrier, Heiter, Masana, Messineo, Mowlavi, Nicolas, Nienartowicz, Pailler, Panuzzo, Riclet, Roux, Seabroke, Sordo, Thévenin, Gracia-Abril, Portell, Teyssier, Altmann, Andrae, Audard, Bellas-Velidis, Benson, Berthier, Blomme, Burgess, Busonero, Busso, Cánovas, Carry, Cellino, Cheek, Clementini, Damerdji, Davidson, de~Teodoro, Nuñez~Campos, Delchambre, Dell'Oro, Esquej, Fernández-Hernández, Fraile, Garabato, García-Lario, Gosset, Haigron, Halbwachs, Hambly, Harrison, Hernández, Hestroffer, Hodgkin, Holl, Janßen, Jevardat~de Fombelle, Jordan, Krone-Martins,
  Lanzafame, Löffler, Marchal, Marrese, Moitinho, Muinonen, Osborne, Pancino, Pauwels, Recio-Blanco, Reylé, Riello, Rimoldini, Roegiers, Rybizki, Sarro, Siopis, Smith, Sozzetti, Utrilla, van Leeuwen, Abbas, Ábrahám, Abreu~Aramburu, Aerts, Aguado, Ajaj, Aldea-Montero, Altavilla, Álvarez, Alves, Anders, Anderson, Anglada~Varela, Antoja, Baines, Baker, Balaguer-Núñez, Balbinot, Balog, Barache, Barbato, Barros, Barstow, Bartolomé, Bassilana, Bauchet, Becciani, Bellazzini, Berihuete, Bernet, Bertone, Bianchi, Binnenfeld, Blanco-Cuaresma, Blazere, Boch, Bombrun, Bossini, Bouquillon, Bragaglia, Bramante, Breedt, Bressan, Brouillet, Brugaletta, Bucciarelli, Burlacu, Butkevich, Buzzi, Caffau, Cancelliere, Cantat-Gaudin, Carballo, Carlucci, Carnerero, Carrasco, Casamiquela, Castellani, Castro-Ginard, Chaoul, Charlot, Chemin, Chiaramida, Chiavassa, Chornay, Comoretto, Contursi, Cooper, Cornez, Cowell, Crifo, Cropper, Crosta, Crowley, Dafonte, Dapergolas, David, David, de~Laverny, De~Luise, De~March, De~Ridder,
  de~Souza, de~Torres, del Peloso, del Pozo, Delbo, Delgado, Delisle, Demouchy, Dharmawardena, Di~Matteo, Diakite, Diener, Distefano, Dolding, Edvardsson, Enke, Fabre, Fabrizio, Faigler, Fedorets, Fernique, Fienga, Figueras, Fournier, Fouron, Fragkoudi, Gai, Garcia-Gutierrez, Garcia-Reinaldos, García-Torres, Garofalo, Gavel, Gavras, Gerlach, Geyer, Giacobbe, Gilmore, Girona, Giuffrida, Gomel, Gomez, González-Núñez, González-Santamaría, González-Vidal, Granvik, Guillout, Guiraud, Gutiérrez-Sánchez, Guy, Hatzidimitriou, Hauser, Haywood, Helmer, Helmi, Sarmiento, Hidalgo, Hilger, Hładczuk, Hobbs, Holland, Huckle, Jardine, Jasniewicz, Jean-Antoine~Piccolo, Jiménez-Arranz, Jorissen, Juaristi~Campillo, Julbe, Karbevska, Kervella, Khanna, Kontizas, Kordopatis, Korn, Kóspál, Kostrzewa-Rutkowska, Kruszyńska, Kun, Laizeau, Lambert, Lanza, Lasne, Le~Campion, Lebreton, Lebzelter, Leccia, Leclerc, Lecoeur-Taibi, Liao, Licata, Lindstrøm, Lister, Livanou, Lobel, Lorca, Loup, Madrero~Pardo, Magdaleno~Romeo,
  Managau, Mann, Manteiga, Marchant, Marconi, Marcos, Marcos~Santos, Marín~Pina, Marinoni, Marocco, Marshall, Martin~Polo, Martín-Fleitas, Marton, Mary, Masip, Massari, Mastrobuono-Battisti, Mazeh, McMillan, Messina, Michalik, Millar, Mints, Molina, Molinaro, Molnár, Monari, Monguió, Montegriffo, Montero, Mor, Mora, Morbidelli, Morel, Morris, Muraveva, Murphy, Musella, Nagy, Noval, Ocaña, Ogden, Ordenovic, Osinde, Pagani, Pagano, Palaversa, Palicio, Pallas-Quintela, Panahi, Payne-Wardenaar, Peñalosa~Esteller, Penttilä, Pichon, Piersimoni, Pineau, Plachy, Plum, Poggio, Prša, Pulone, Racero, Ragaini, Rainer, Raiteri, Rambaux, Ramos, Ramos-Lerate, Re~Fiorentin, Regibo, Richards, Rios~Diaz, Ripepi, Riva, Rix, Rixon, Robichon, Robin, Robin, Roelens, Rogues, Rohrbasser, Romero-Gómez, Rowell, Royer, Ruz~Mieres, Rybicki, Sadowski, Sáez~Núñez, Sagristà~Sellés, Sahlmann, Salguero, Samaras, Sanchez~Gimenez, Sanna, Santoveña, Sarasso, Schultheis, Sciacca, Segol, Segovia, Ségransan, Semeux, Shahaf,
  Siddiqui, Siebert, Siltala, Silvelo, Slezak, Slezak, Smart, Snaith, Solano, Solitro, Souami, Souchay, Spagna, Spina, Spoto, Steele, Steidelmüller, Stephenson, Süveges, Surdej, Szabados, Szegedi-Elek, Taris, Taylor, Teixeira, Tolomei, Tonello, Torra, Torra, Torralba~Elipe, Trabucchi, Tsounis, Turon, Ulla, Unger, Vaillant, van Dillen, van Reeven, Vanel, Vecchiato, Viala, Vicente, Voutsinas, Weiler, Wevers, Wyrzykowski, Yoldas, Yvard, Zhao, Zorec, Zucker, \& Zwitter}]{gaia_collaboration_gaia_2023}
{Gaia Collaboration}, Vallenari, A., Brown, A. G.~A., {et~al.} 2023, \aap, 674, A1

\bibitem[{Gontcharov(2017)}]{gontcharov_3d_2017}
Gontcharov, G.~A. 2017, Astronomy Letters, 43, 472

\bibitem[{Gustafsson {et~al.}(2008)Gustafsson, Edvardsson, Eriksson, Jørgensen, Nordlund, \& Plez}]{gustafsson_grid_2008}
Gustafsson, B., Edvardsson, B., Eriksson, K., {et~al.} 2008, \aap, 486, 951

\bibitem[{Höfner \& Olofsson(2018)}]{hofner_mass_2018}
Höfner, S. \& Olofsson, H. 2018, \aapr, 26, 1

\bibitem[{Lenz \& Breger(2005)}]{lenz_period04_2005}
Lenz, P. \& Breger, M. 2005, Communications in Asteroseismology, 146, 53

\bibitem[{Lindegren {et~al.}(2021)Lindegren, Bastian, Biermann, Bombrun, de~Torres, Gerlach, Geyer, Hernández, Hilger, Hobbs, Klioner, Lammers, McMillan, Ramos-Lerate, Steidelmüller, Stephenson, \& van Leeuwen}]{lindegren_gaia_2021}
Lindegren, L., Bastian, U., Biermann, M., {et~al.} 2021, \aap, 649, A4

\bibitem[{McDonald \& Zijlstra(2015)}]{mcdonald_mass-loss_2015}
McDonald, I. \& Zijlstra, A.~A. 2015, \mnras, 448, 502

\bibitem[{Merchan-Benitez {et~al.}(2023)Merchan-Benitez, Uttenthaler, \& Jurado-Vargas}]{merchan-benitez_meandering_2023}
Merchan-Benitez, P., Uttenthaler, S., \& Jurado-Vargas, M. 2023, \aap, 672, A165

\bibitem[{Pojmanski(1998)}]{pojmanski_all_1998}
Pojmanski, G. 1998, Acta Astronomica, 48, 35

\bibitem[{Templeton {et~al.}(2005)Templeton, Mattei, \& Willson}]{templeton_secular_2005}
Templeton, M.~R., Mattei, J.~A., \& Willson, L.~A. 2005, \aj, 130, 776

\bibitem[{Trabucchi {et~al.}(2019)Trabucchi, Wood, Montalbán, Marigo, Pastorelli, \& Girardi}]{trabucchi_modelling_2019}
Trabucchi, M., Wood, P.~R., Montalbán, J., {et~al.} 2019, \mnras, 482, 929

\bibitem[{Uttenthaler {et~al.}(2019)Uttenthaler, McDonald, Bernhard, Cristallo, \& Gobrecht}]{uttenthaler_interplay_2019}
Uttenthaler, S., McDonald, I., Bernhard, K., Cristallo, S., \& Gobrecht, D. 2019, \aap, 622, A120

\bibitem[{Vassiliadis \& Wood(1993)}]{vassiliadis_evolution_1993}
Vassiliadis, E. \& Wood, P.~R. 1993, \apj, 413, 641

\bibitem[{Wittkowski {et~al.}(2018)Wittkowski, Rau, Chiavassa, Höfner, Scholz, Wood, de~Wit, Eisenhauer, Haubois, \& Paumard}]{wittkowski_vlti-gravity_2018}
Wittkowski, M., Rau, G., Chiavassa, A., {et~al.} 2018, \aap, 613, L7

\bibitem[{Wood \& Zarro(1981)}]{wood_helium-shell_1981}
Wood, P.~R. \& Zarro, D.~M. 1981, \apj, 247, 247

\end{thebibliography}

\begin{appendix} 
\section{Tables}
\begin{table*}[ht!]
\small
\begin{center}
 \caption{RHD simulation parameters.}
 \label{tab_simus}
 \begin{tabular}{l|rrrrrrrrr|rrrr}
\hline
\multicolumn{1}{c|}{Id} & \multicolumn{1}{c}{Simulation} & \multicolumn{1}{c}{$M_\star$} & \multicolumn{1}{c}{$L_\star$} & \multicolumn{1}{c}{$R_\star$} & \multicolumn{1}{c}{$\mathrm{T_{eff}}$} & \multicolumn{1}{c}{$\log g$} & \multicolumn{1}{c}{$\mathrm{t_{avg}}$} & \multicolumn{1}{c}{$\mathrm{P_{puls}}$} & \multicolumn{1}{c}{$\mathrm{\sigma_{Ppuls}}$} & \multicolumn{1}{c}{$\langle P_x\rangle$} & \multicolumn{1}{c}{$\langle P_y\rangle$} & \multicolumn{1}{c}{$\mathrm{\sigma_P}$} & \multicolumn{1}{c}{$\mathrm{\sigma_P}$}\\
\multicolumn{1}{c|}{} & \multicolumn{1}{c}{} & \multicolumn{1}{c}{[$M_\odot$]} & \multicolumn{1}{c}{[$L_\odot$]} & \multicolumn{1}{c}{[$R_\odot$]} & \multicolumn{1}{c}{[K]} & \multicolumn{1}{c}{[cgs]} & \multicolumn{1}{c}{[yr]} & \multicolumn{1}{c}{[days]} & \multicolumn{1}{c}{[days]} & \multicolumn{1}{c}{[AU]} & \multicolumn{1}{c}{[AU]} & \multicolumn{1}{c}{[AU]} & \multicolumn{1}{c}{\% [$R_\star$]}\\
\hline
1 & st28gm05n038 & 1.0 & 4978 & 281 & 2893 & -0.46 & 19.05 & 341 & 31 & -0.025 & -0.004 & 0.108 & 8.26\\
2 & st28gm05n043 & 1.0 & 5013 & 278 & 2908 & -0.45 & 19.03 & 337 & 30 & -0.011 & -0.044 & 0.113 & 8.74\\
3 & st28gm06n057 & 1.0 & 7055 & 349 & 2831 & -0.65 & 15.87 & 485 & 77 & 0.012 & 0.005 & 0.152 & 9.37 \\
4 & st28gm06n059 & 1.0 & 7050 & 347 & 2837 & -0.64 & 15.87 & 487 & 77 & 0.082 & -0.055 & 0.143 & 8.86\\
5 & st28gm07n006 & 1.0 & 8003 & 379 & 2801 & -0.72 & 15.85 & 546 & 99 & -0.055 & 0.011 & 0.219 & 12.43\\
6 & st28gm07n008 & 1.0 & 8020 & 377 & 2812 & -0.72 & 15.84 & 541 & 102 & 0.006 & -0.029 & 0.171 & 9.75 \\
7 & st28gm08n001 & 1.0 & 8980 & 408 & 2780 & -0.78 & 19.39 & 592 & 129 & -0.141 & -0.018 & 0.202 & 10.65\\
8 & st29gm02n013 & 1.0 & 2983 & 190 & 3092 & -0.12 & 11.23 & 178 & 13 & -0.015 & -0.012 & 0.052 & 5.89 \\
9 & st29gm03n001 & 1.0 & 3990 & 241 & 2955 & -0.33 & 15.22 & 262 & 18 & 0.001 & -0.001 & 0.094 & 8.39 \\
10 & st29gm03n002 & 1.0 & 4028 & 241 & 2964 & -0.33 & 15.22 & 263 & 18 & 0.008 & -0.008 & 0.133 & 11.87\\
11 & st31gm01n002 & 1.0 & 2500 & 165 & 3177 & 0.01 & 9.13 & 134  & 14 & -0.002 & -0.003 & 0.028 & 3.65\\
12 & st32g01n002 & 1.0 & 1978 & 138 & 3275 & 0.16 & 7.02 & 98 & 8 & -0.006 & -0.007 & 0.025 & 3.90\\ 
\hline
13 & st28gm05n006 & 1.5 & 4985 & 269 & 2957 & -0.25 & 15.85 & 240  & 20 & -0.003 & 0.004 & 0.087 & 6.95\\
14 & st28gm05n008 & 1.0 & 4942 & 302 & 2786 & -0.52 & 15.85 & 387 & 36 &  0.003 & -0.032 & 0.067 & 4.77\\
15 & st28gm05n017 & 1.5 & 7490 & 321 & 2994 & -0.40 & 15.86 & 328 & 22 & -0.013 & -0.072 & 0.129 & 8.64\\ 
16 & st28gm05n020 & 1.5 & 7708 & 327 & 2989 & -0.42 & 15.86 & 351 & 30 & -0.046 & -0.023 & 0.130 & 8.55\\
17 & st28gm05n022 & 1.0 & 5068 & 312 & 2757 & -0.55 & 15.85 & 422 & 42 & 0.005 & -0.027 & 0.086 & 5.93 \\
18 & st28gm05n028 & 1.5 & 6946 & 306 & 3009 & -0.36 & 15.85 & 298 & 16 & 0.028 & -0.015 & 0.104 & 7.31 \\ 
19 & st28gm05n029 & 1.5 & 6941 & 288 & 3102 & -0.31 & 15.70 & 278 & 15 & 0.024 & -0.010 & 0.093 & 6.94 \\
20 & st28gm05n034 & 1.5 & 6880 & 278 & 3150 & -0.28 & 7.93 & 259 & 14 & -0.004 & -0.040 & 0.082 & 6.34 \\
21 & st28gm06n032 & 1.0 & 7062 & 339 & 2872 & -0.62 & 15.87 & 508 & 69 & -0.034 & 0.011 & 0.156 & 9.90\\
22 & st28gm06n043 & 1.0 & 7079 & 344 & 2854 & -0.64 & 15.87 & 461 & 75 & -0.062 & -0.041 & 0.108 & 6.75\\
23 & st28gm06n053 & 1.5 & 10073 & 391 & 2922 & -0.57 & 15.85 & 418 & 46 & 0.009 & -0.068 & 0.238 & 13.09\\
\hline
24 & st26gm07n002 & 1.0 & 6986 & 439 & 2524 & -0.85 & 25.35 & 594 & 112 & -0.100\tablefootmark{a} & 0.046\tablefootmark{a} & 0.187\tablefootmark{a} & 9.16\tablefootmark{a} \\
25 & st26gm07n001 & 1.0 & 6953 & 402 & 2635 & -0.77 & 27.74 & 517 & 94 & -0.098\tablefootmark{a} & 0.024\tablefootmark{a} & 0.198\tablefootmark{a} & 10.59\tablefootmark{a} \\
26 & st28gm06n026 & 1.0 & 6955 & 372 & 2737 & -0.70 & 25.35 & 471 & 116 & -0.068\tablefootmark{a} & -0.002\tablefootmark{a} & 0.152\tablefootmark{a} & 8.79\tablefootmark{a} \\
27 & st29gm06n001 & 1.0 & 6995 & 324 & 2929 & -0.59 & 31.70 & 389 & 73 & -0.098\tablefootmark{a} & 0.016\tablefootmark{a} & 0.174\tablefootmark{a} & 11.55\tablefootmark{a} \\
28 & st27gm06n001 & 1.0 & 5011 & 322 & 2704 & -0.58 & 31.73 & 450 & 38 & -0.027 & 0.027 & 0.090 & 6.01 \\
29 & st28gm05n002 & 1.0 & 4978 & 314 & 2742 & -0.56 & 25.35 & 393 & 38 & -0.002\tablefootmark{a} & 0.033\tablefootmark{a} & 0.077\tablefootmark{a} & 5.27\tablefootmark{a} \\
30 & st28gm05n001 & 1.0 & 5019 & 289 & 2858 & -0.49 & 31.83 & 360 & 44 & -0.057 & 0.017 & 0.097 & 7.22\\
31 & st29gm04n001 & 1.0 & 4982 & 295 & 2827 & -0.50 & 25.35 & 339 & 37 & -0.002\tablefootmark{a} & 0.023\tablefootmark{a} & 0.078\tablefootmark{a} & 5.69\tablefootmark{a}\\
 \end{tabular}
\end{center}
{\textbf{Notes:} The simulations 1-12 are the new models presented in this work; the simulations 13-23 are presented in \cite{ahmad_properties_2023} and the simulations 24-31 are presented in \cite{freytag_global_2017} and \cite{chiavassa_heading_2018}. 
The table displays the simulation name, the stellar mass $\mathrm{M_\star}$, the average emitted luminosity $\mathrm{L_\star}$, the average approximate stellar radius $\mathrm{R_\star}$, the effective temperature $\mathrm{T_{eff}}$, the surface gravity $ \mathrm{\log(g)}$, the pulsation period $\mathrm{P_{puls}}$, the spread in the pulsation period i.e. the pulsation period uncertainty $\mathrm{\sigma_{puls}}$, and the stellar time $\mathrm{t_{avg}}$ used for the averaging of the rest of the quantities. The stellar parameters may slightly vary from original articles as they are updated thanks to Ahmad's and Freytag's work. The last four columns are the time-averaged positions $P_x$ and $P_y$ in AU and the standard deviation of the photocentre displacement $\mathrm{\sigma_P}$ (in AU and in $\%$ of $\mathrm{R_\star}$). Data denoted by the footnote \tablefoottext{a} come from the previous analysis of \cite{chiavassa_heading_2018}.}
\end{table*}

\begin{table*}
\small
 \caption{Parameters of the sample Miras.}
\begin{center}
 \label{Table_Miras}
 \begin{tabular}{l|ccccccccccc}
\hline
Name & $\mathrm{P_{obs}}$ & RUWE & $\mathrm{N_{per}}$ & $\mathrm{N_{good}}$ & $L_\star$ & $L_\star^-$ & $L_\star^+$ & $\varpi$ & $\varpi_\mathrm{{corr}}$ & $\mathrm{\sigma_\varpi}$ & Population  \\
 & [days] &  &  & & [$L_\odot$] & [$L_\odot$] & [$L_\odot$] & [mas] & [mas] & [mas] & \\
 \hline

Y And & 221 & 0.774 & 16 & 297 & 4112 & 431 & 530 & 0.557 & 0.613 & 0.062 & thin\\
RT Aql & 328 & 1.003 & 17 & 277 & 5623 & 632 & 703 & 1.793 & 1.844 & 0.102 & thin\\
SY Aql & 356 & 1.367 & 22 & 632 & 5623 & 1045 & 1253 & 1.067 & 1.121 & 0.091 & thick\\
V335 Aql & 176 & 0.931 & 18 & 322 &  2813 & 561 & 917 & 0.191 & 0.219 & 0.047 & halo\\
T Aqr & 202 & 0.832 & 21 & 479 & 3317 & 255 & 275 & 0.906 & 0.960 & 0.043 & thin\\
U Ari & 372 & 1.240 & 13 & 131 & 4402 & 381 & 420 & 1.674 & 1.729 & 0.110 & thin\\
RU Aur & 470 & 1.348 & 16 & 230 & 5125 & 1387 & 1847 & 1.343 & 1.400 & 0.126 & thin\\
R Boo & 224 & 1.049 & 20 & 314 & 4255 & 821 & 1009 & 1.520 & 1.568 & 0.059 & thick\\
T Cap & 271 & 0.883 & 12 & 253 & 7124 & 985 & 1310 & 0.566 & 0.624 & 0.086 & thick\\
CM Car & 339 & 0.955 & 27 & 343 & 9404 & 1538 & 2271 & 0.275 & 0.294 & 0.053 & thick\\
U Cet & 234 & 1.011 & 16 & 283 & 4631 & 764 & 897 & 0.954 & 1.000 & 0.071 & thick\\
R Cha & 338 & 1.375 & 24 & 322 & 4574 & 999 & 1258 & 1.076 & 1.102 & 0.062 & thick\\
S CMi & 334 & 1.138 & 13 & 168 & 6174 & 378 & 402 & 2.393 & 2.440 & 0.098 & thin\\
U CMi & 410 & 0.959 & 15 & 252 & 9342 & 888 & 1064 & 0.672 & 0.723 & 0.066 & thin\\
W Cnc & 394 & 1.336 & 15 & 252 & 5966 & 512 & 571 & 1.897 & 1.935 & 0.134 & thin\\
R Col & 328 & 0.927 & 28 & 452 & 7233 & 1611 & 2031 & 0.696 & 0.730 & 0.049 & thin\\
T Col & 226 & 0.998 & 24 & 338 & 3432 & 211 & 224 & 1.603 & 1.628 & 0.041 & thick\\
RY CrA & 206 & 1.144 & 19 & 302 & 1819 & 455 & 899 & 0.179 & 0.193 & 0.059 & thick\\
X CrB & 241 & 0.934 & 30 & 468 & 4776 & 1052 & 1324 & 0.661 & 0.708 & 0.043 & thick\\
R Del & 286 & 0.907 & 21 & 430 & 5144 & 595 & 664 & 1.265 & 1.318 & 0.064 & thick\\
W Dra & 290 & 0.897 & 25 & 381 & 7074 & 980 & 1350 & 0.213 & 0.257 & 0.034 & halo\\
T Eri & 252 & 1.054 & 24 & 690 & 4099 & 278 & 306 & 1.109 & 1.146 & 0.065 & thick\\
U Eri & 274 & 0.890 & 27 & 761 & 4581 & 433 & 522 & 0.516 & 0.558 & 0.051 & thin\\
V Gem & 275 & 1.299 & 14 & 183 & 3123 & 357 & 421 & 1.070 & 1.125 & 0.110 & thin\\
S Her & 304 & 1.122 & 18 & 644 & 5531 & 434 & 466 & 1.712 & 1.756 & 0.059 & thin\\
SV Her & 238 & 0.815 & 27 & 531 & 6339 & 807 & 1065 & 0.306 & 0.359 & 0.044 & thick\\
T Her & 164 & 0.923 & 26 & 379 & 2174 & 195 & 212 & 1.153 & 1.198 & 0.041 & thick\\
T Hor & 219 & 0.962 & 28 & 390 & 2886 & 200 & 217 & 0.904 & 0.933 & 0.047 & thick\\
RR Hya & 342 & 0.811 & 16 & 199 & 5887 & 528 & 638 & 0.710 & 0.753 & 0.069 & thin\\
RU Hya & 333 & 1.274 & 14 & 176 & 7538 & 559 & 628 & 1.208 & 1.261 & 0.084 & thick\\
S Lac & 240 & 1.188 & 24 & 870 & 3330 & 717 & 905 & 1.254 & 1.301 & 0.052 & thin\\
RR Lib & 278 & 1.145 & 16 & 329 & 4629 & 402 & 449 & 1.007 & 1.062 & 0.073 & thin\\
R LMi & 373 & 1.106 & 17 & 402 & 5288 & 327 & 348 & 3.446 & 3.496 & 0.142 & thin\\
RT Lyn & 394 & 0.761 & 19 & 277 & 7182 & 1740 & 2237 & 0.665 & 0.722 & 0.058 & thick\\
U Oct & 303 & 1.034 & 27 & 329 & 5868 & 518 & 563 & 1.009 & 1.035 & 0.052 & thin\\
R Oph & 303 & 1.276 & 16 & 251 & 5079 & 431 & 470 & 1.886 & 1.938 & 0.107 & thick\\
RY Oph & 151 & 0.829 & 17 & 418 &  1908 & 129 & 138 & 1.324& 1.377 & 0.056 & thick\\
SY Pav & 191 & 0.866 & 19 & 338 & 1237 & 217 & 264 & 0.349 & 0.348 & 0.045 & thick\\
R Peg & 378 & 1.276 & 11 & 131 & 4244 & 336 & 363 & 2.629 & 2.681 & 0.117 & thin\\
S Peg & 314 & 1.017 & 13 & 175 & 6545 & 531 & 583 & 1.345 & 1.399 & 0.083 & thick\\
X Peg & 201 & 0.625 & 19 & 290 & 5913 & 673 & 783 & 0.428 & 0.483 & 0.042 & thick\\
Z Peg & 328 & 1.010 & 18 & 643 & 8175 & 545 & 581 & 1.520 & 1.571 & 0.060 & thin\\
RZ Sco & 159 & 1.279 & 16 & 511 &  3173 & 333 & 391 & 0.642 & 0.679 & 0.063 & halo\\
R Sgr & 269 & 1.146 & 14 & 157 & 6466 & 610 & 680 & 1.138 & 1.190 & 0.081 & thin\\
RV Sgr & 319 & 0.830 & 15 & 187 & 6195 & 383 & 415 & 1.306 & 1.357 & 0.067 & thin\\
BH Tel & 217 & 0.912 & 18 & 311 & 3417 & 736 & 1269 & 0.220 & 0.245 & 0.060 & thin\\
T Tuc & 247 & 1.065 & 32 & 509 & 3435 & 401 & 448 & 0.876 & 0.911 & 0.045 & thin\\
RR UMa & 231 & 0.861 & 29 & 398 & 3333 & 356 & 450 & 0.351 & 0.389 & 0.042 & thick\\
T UMa & 256 & 1.289 & 25 & 378 & 3963 & 1103 & 1503 & 0.989 & 1.019 & 0.065 & thick\\
T UMi & 235 & 1.363 & 26 & 355 & 4821 & 1258 & 1673 & 0.784 & 0.810 & 0.050 & thick\\
CI Vel & 138 & 1.121 & 23 & 372 & 3112 & 457 & 571 & 0.223 & 0.261 & 0.031 & thick\\
R Vir & 146 & 0.856 & 15 & 383 & 1811 & 129 & 138 & 2.196 & 2.248 & 0.050 & thin\\
R Vul & 137 & 0.805 & 18 & 340 & 1292 & 147 & 164 & 1.437 & 1.488 & 0.047 & thin\\
\end{tabular}
\end{center}
{\textbf{Notes:} The columns are:  Miras' name; the pulsation period $\mathrm{P_{obs}}$ obtained from light curves, its uncertainty is assumed to be $2.4\%$ of the corresponding $\mathrm{P_{obs}}$ \citep{merchan-benitez_meandering_2023}; RUWE; $\mathrm{N_{per}}$ the number of visibility periods used in the astrometric solution\footnote{}, a visibility period consists of a group of observations separated from other groups by at least $4$ days. A high number of periods is a indicator of a well-observed source while a value smaller than $10$ indicates that the calculated parallax could be more vulnerable to errors (\textit{visibility\_periods\_used} in the \textit{Gaia} archive); $\mathrm{N_{good}}$ the total number of good observations along-scan (\textit{astrometric\_n\_good\_obs\_al}) by \textit{Gaia} to compute the astrometric solution; the luminosity $\mathrm{L_\star}$; the negative luminosity uncertainty $\mathrm{L_\star^-}$; the positive luminosity uncertainty $\mathrm{L_\star^+}$; the GDR3 parallax $\varpi$; the corrected GDR3 parallax $\varpi_{\mathrm{corr}}$ according to \cite{lindegren_gaia_2021}; the parallax uncertainty $\mathrm{\sigma_\varpi}$; population membership based on a study of stellar total space velocity according to \cite{chen_planets_2021}, halo stars are more metal poor.}
\end{table*}
\clearpage

\begin{table*}
\small
 \caption{Stellar parameters of the Miras inferred from the simulations.}
\begin{center}
 \label{Table_Miras_RES}
 \begin{tabular}{l|cccccccccc}
\hline
Name & $\mathrm{P_{1.0}}$ & $\mathrm{\Delta P_{1.0}}$ & $\mathrm{P_{1.5}}$ & $\mathrm{\Delta P_{1.5}}$ & log($g_{1.0}$) & log($g_{1.5}$) & $\mathrm{R_{1.0}}$ & $\mathrm{R_{1.5}}$ & $\mathrm{T_{1.0}}$ & $\mathrm{T_{1.5}}$\\
 & [days] & [\%] & [days] & [\%] & [cgs] & [cgs] & [$R_\odot$] & [$R_\odot$] & [K] & [K]\\
\hline
Y And & 270 & 22 & 254 & 15 & -0.35 & -0.27 & 246 & 274 & 3012 & 3042\\
RT Aql & 380 & 16 & 367 & 12 & -0.52 & -0.47 & 302 & 348 & 2843 & 2860\\
SY Aql & 353 & 1 & 339 & 5 & -0.48 & -0.43 & 289 & 330 & 2879 & 2898\\
V335 Aql & 225 & 28 & 209 & 19 & -0.25 & -0.16 & 221 & 242 & 3106 & 3144\\
T Aqr & 213 & 5 & 197 & 2 & -0.22 & -0.13 & 214 & 232 & 3136 & 3177\\
U Ari & 402 & 8 & 390 & 5 & -0.55 & -0.50 & 312 & 362 & 2816 & 2830\\
RU Aur & 443 & 6 & 434 & 8 & -0.60 & -0.56 & 331 & 387 & 2770 & 2780\\
R Boo & 261 & 17 & 245 & 10 & -0.33 & -0.25 & 241 & 268 & 3029 & 3061\\
T Cap & 337 & 25 & 323 & 19 & -0.46 & -0.40 & 281 & 320 & 2901 & 2922\\
CM Car & 246 & 28 & 230 & 32 & -0.30 & -0.21 & 233 & 257 & 3060 & 3095\\
U Cet & 298 & 27 & 283 & 21 & -0.40 & -0.33 & 261 & 293 & 2962 & 2989\\
R Cha & 270 & 20 & 254 & 25 & -0.35 & -0.27 & 246 & 274 & 3012 & 3042\\
S CMi & 371 & 11 & 358 & 7 & -0.51 & -0.46 & 298 & 342 & 2854 & 2872\\
U CMi & 283 & 31 & 267 & 35 & -0.37 & -0.29 & 253 & 283 & 2988 & 3017\\
W Cnc & 464 & 18 & 456 & 16 & -0.63 & -0.59 & 340 & 400 & 2748 & 2757\\
R Col & 231 & 30 & 215 & 35 & -0.27 & -0.17 & 224 & 246 & 3093 & 3130\\
T Col & 207 & 9 & 191 & 16 & -0.21 & -0.11 & 210 & 228 & 3151 & 3194\\
RY CrA & 262 & 27 & 247 & 20 & -0.33 & -0.25 & 242 & 269 & 3026 & 3058\\
X CrB & 215 & 11 & 199 & 17 & -0.23 & -0.13 & 215 & 234 & 3130 & 3171\\
R Del & 277 & 3 & 261 & 9 & -0.36 & -0.28 & 250 & 279 & 2999 & 3028\\
W Dra & 184 & 36 & 169 & 42 & -0.15 & -0.04 & 196 & 210 & 3212 & 3260\\
T Eri & 281 & 11 & 265 & 5 & -0.37 & -0.29 & 252 & 282 & 2992 & 3021\\
U Eri & 239 & 13 & 223 & 19 & -0.28 & -0.19 & 229 & 252 & 3075 & 3111\\
V Gem & 402 & 46 & 391 & 42 & -0.55 & -0.5 & 312 & 362 & 2815 & 2829\\
S Her & 261 & 14 & 245 & 19 & -0.33 & -0.25 & 242 & 268 & 3028 & 3061\\
SV Her & 215 & 10 & 199 & 16 & -0.23 & -0.13 & 215 & 234 & 3130 & 3170\\
T Her & 206 & 26 & 190 & 16 & -0.21 & -0.11 & 210 & 227 & 3153 & 3195\\
T Hor & 227 & 4 & 211 & 4 & -0.26 & -0.16 & 222 & 243 & 3102 & 3140\\
RR Hya & 290 & 15 & 274 & 20 & -0.38 & -0.31 & 257 & 288 & 2976 & 3004\\
RU Hya & 332 & 1 & 317 & 5 & -0.45 & -0.39 & 279 & 316 & 2908 & 2930\\
S Lac & 242 & 1 & 226 & 6 & -0.29 & -0.20 & 231 & 254 & 3067 & 3103\\
RR Lib & 302 & 9 & 287 & 3 & -0.4 & -0.33 & 263 & 296 & 2956 & 2982\\
R LMi & 482 & 29 & 475 & 27 & -0.65 & -0.61 & 348 & 411 & 2730 & 2737\\
RT Lyn & 258 & 34 & 243 & 38 & -0.32 & -0.24 & 240 & 266 & 3034 & 3067\\
U Oct & 240 & 21 & 224 & 26 & -0.29 & -0.20 & 230 & 253 & 3072 & 3108\\
R Oph & 393 & 30 & 381 & 26 & -0.54 & -0.49 & 308 & 356 & 2827 & 2842\\
RY Oph & 252 & 67 & 236 & 57 & -0.31 & -0.23 & 236 & 261 & 3047 & 3081\\
SY Pav & 219 & 15 & 203 & 6 & -0.24 & -0.14 & 217 & 237 & 3121 & 3161\\
R Peg & 421 & 11 & 410 & 9 & -0.58 & -0.53 & 321 & 373 & 2794 & 2806\\
S Peg & 331 & 5 & 316 & 1 & -0.45 & -0.39 & 278 & 316 & 2910 & 2932\\
X Peg & 209 & 4 & 194 & 4 & -0.22 & -0.12 & 212 & 230 & 3144 & 3186\\
Z Peg & 264 & 19 & 248 & 24 & -0.33 & -0.25 & 243 & 270 & 3023 & 3055\\
RZ Sco & 274 & 72 & 258 & 62 & -0.35 & -0.28 & 249 & 277 & 3004 & 3034\\
R Sgr & 326 & 21 & 311 & 16 & -0.44 & -0.38 & 275 & 312 & 2918 & 2941\\
RV Sgr & 286 & 10 & 270 & 15 & -0.38 & -0.30 & 255 & 285 & 2983 & 3012\\
BH Tel & 264 & 22 & 248 & 14 & -0.33 & -0.25 & 243 & 270 & 3024 & 3056\\
T Tuc & 219 & 11 & 203 & 18 & -0.24 & -0.14 & 218 & 237 & 3120 & 3159\\
T UMa & 279 & 9 & 263 & 3 & -0.36 & -0.29 & 251 & 280 & 2996 & 3025\\
RR UMa & 209 & 10 & 193 & 16 & -0.21 & -0.12 & 212 & 230 & 3145 & 3186\\
T UMi & 236 & 1 & 220 & 6 & -0.28 & -0.19 & 227 & 250 & 3081 & 3118\\
CI Vel & 173 & 25 & 158 & 14 & -0.12 & 0.00 & 189 & 201 & 3247 & 3297\\
R Vir & 236 & 62 & 220 & 51 & -0.28 & -0.19 & 227 & 250 & 3081 & 3118\\
R Vul & 226 & 66 & 211 & 54 & -0.26 & -0.16 & 222 & 243 & 3103 & 3141\\
\end{tabular}
\end{center}
{\textbf{Notes:} The subscripts $1.0$ denotes quantities derived from the $\mathrm{1.0\, M_\odot}$ simulations, and $1.5$ those from the $\mathrm{1.5\, M_\odot}$ simulations. The columns are: Miras' name; the pulsation periods $P_{1.0}$ and $\mathrm{P_{1.5}}$ in days; the relative difference between the observed pulsation period and our results $\mathrm{\Delta P_{1.0}}$ and $\mathrm{\Delta P_{1.5}}$ in \%; the effective surface gravity $\mathrm{\log(g_{1.0})}$ and $\mathrm{\log(g_{1.5})}$, with $g$ in cgs; $\mathrm{R_{1.0}}$ and $\mathrm{R_{1.5}}$ the radius in $R_\odot$; $\mathrm{T_{1.0}}$ and $\mathrm{T_{1.5}}$ the effective temperature in $K$.}
\end{table*}
\clearpage

\renewcommand{\thefigure}{C\arabic{figure}}
\setcounter{figure}{0}  
\section*{Appendix C: M dwarfs versus AGB stars parallax uncertainty}

Our key assumption is that the parallax uncertainty budget in the Mira sample is dominated by the photocentre shift due to the huge AGB convection cells. To test this assumption, one would need a comparison sample of stars with similar properties such as apparent $G$ magnitude, distance, $\mathrm{G_{BP}-G_{RP}}$ colour, etc., but ideally without surface brightness inhomogeneities. M-type dwarfs could be useful for a comparison because they have similar $\mathrm{G_{BP}-G_{RP}}$ colour as our Miras. Therefore, we searched the SIMBAD database for M5 dwarfs with $G<15$\,mag, which yielded a sample of 240 objects. The list was cross-matched with the $Gaia$ DR3 catalogue. Obvious misidentifications between SIMBAD and $Gaia$ with $G>15$ were culled from the list. A Hertzsprung-Russell diagram based on $\mathrm{M_G}$ vs. $\mathrm{G_{BP}-G_{RP}}$ revealed that the sample still contained several misclassified M-type giant stars. Removing them retained a sample of 99 dwarf stars that have comparable $\mathrm{G_{BP}-G_{RP}}$ colour to the Miras sample. However, as M dwarfs are intrinsically much fainter than Miras, the dwarf stars are much closer to the sun than the Miras: their distances vary between $\sim7$ and 160\,pc, whereas our Mira sample stars are located between 300 and over 5000\,pc from the sun. Furthermore, we noticed that a significant fraction of the dwarfs have surprisingly large parallax uncertainties. These could be related to strong magnetic fields on the surfaces of these dwarfs that are the cause of bright flares or large, dark spots, creating surface brightness variations similar to those expected in the AGB stars. A detailed investigations into the reasons for their large parallax uncertainties is beyond the scope of this paper. We therefore decided to not do the comparison with the M dwarfs.

\begin{figure}[h!]
    \centering
    \includegraphics[width=\columnwidth]{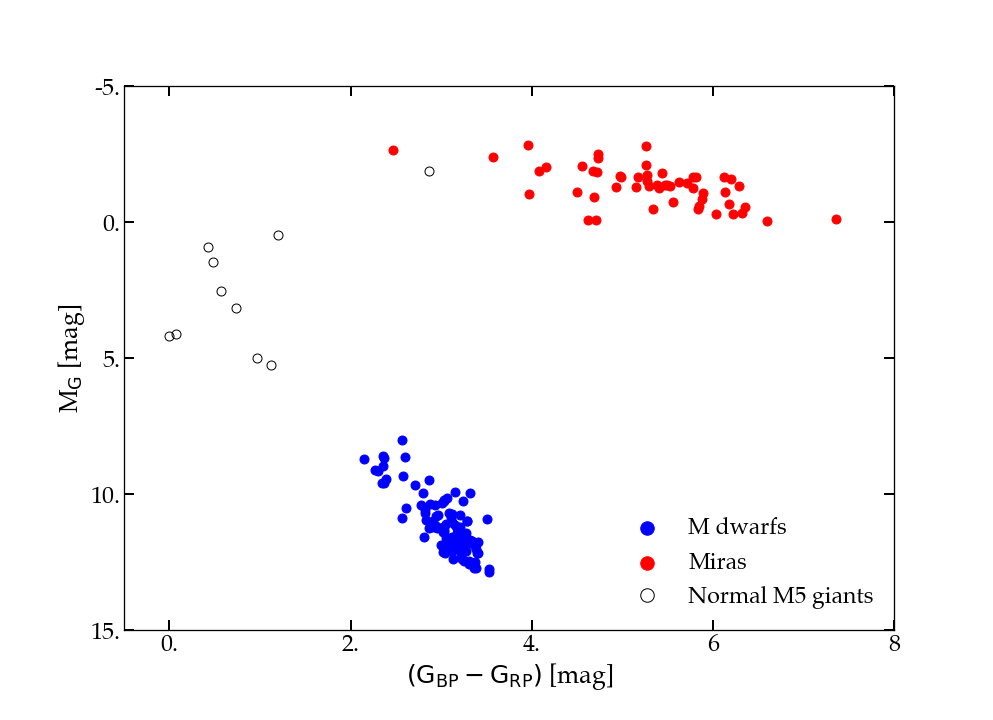}
    \caption{Hertzsprung-Russell diagram: absolute $G$ magnitude $\mathrm{M_G}$ versus $\mathrm{G_{BP}-G_{RP}}$ colour from $Gaia$ data. In red are the Mira stars of our sample, in blue the M5 dwarfs, and in white the misclassified (normal) M giants.} 
    \label{app_Mdwarf_HR}
\end{figure}

Luckily, the contaminant, misclassified (normal) M giants in the Simbad search appear to be a much better comparison sample. They have overlap with the Mira stars in $G$ magnitude and are at fairly similar distances, between $\sim260$ and 1700\,pc. The only drawbacks are that the normal M giants are somewhat bluer in $\mathrm{G_{BP}-G_{RP}}$ colour than the Miras, and we found only ten suitable M giants in our limited search. Fig. \ref{app_Mdwarf_HR} illustrates the location of the M giants together with the Mira sample and the M dwarfs in an HR diagram.

\begin{figure}[h!]
    \centering
    \includegraphics[width=\columnwidth]{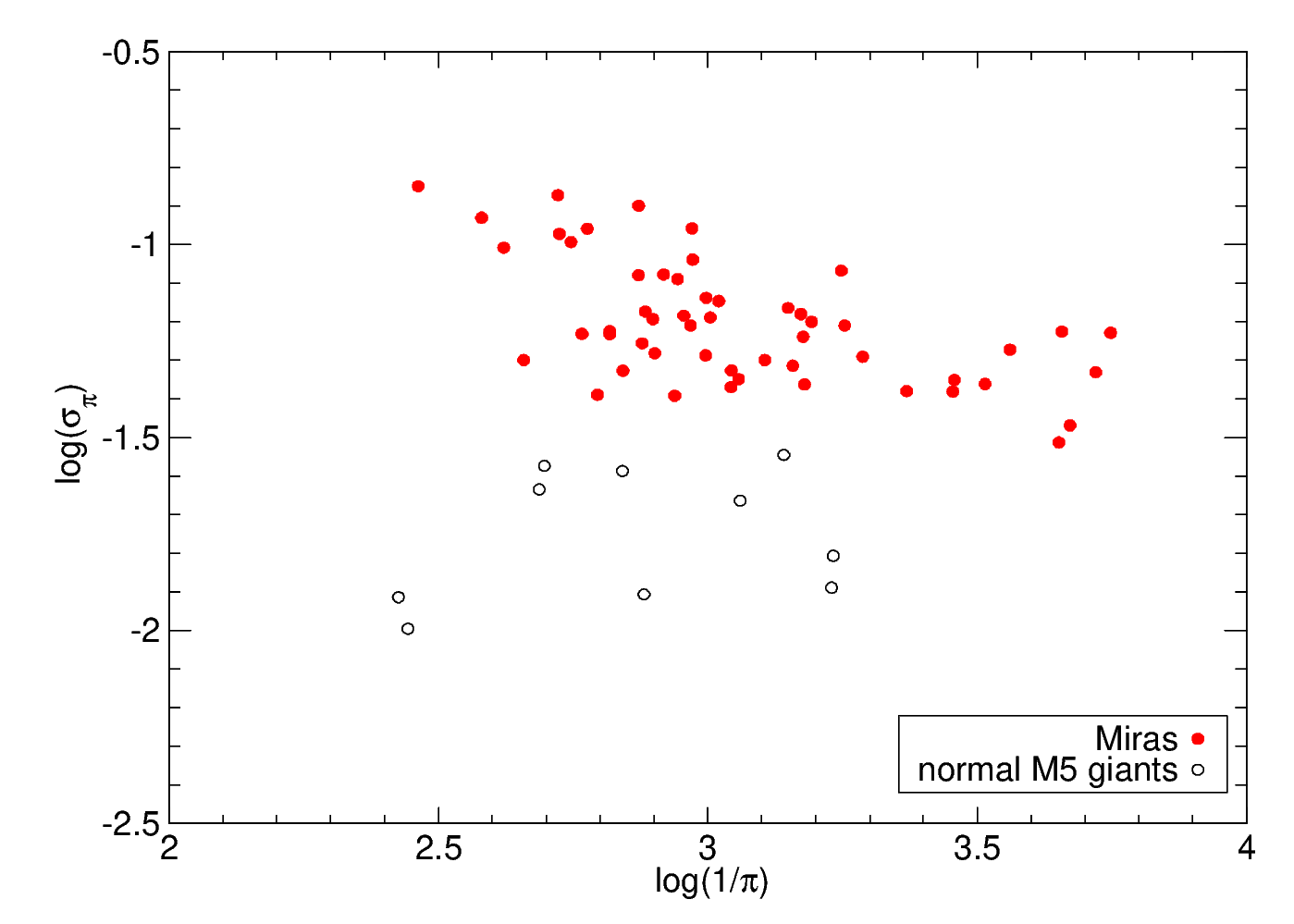}
    \caption{Log-log of the parallax uncertainty, $\mathrm{\sigma_\varpi}$, versus the distance, simply taken as the inverse of the parallax, $\varpi$. We see that the M giants parallax uncertainties are smaller than those of the Miras by a factor of $\sim3.5$.}
    \label{app_Mdwarf}
\end{figure}

Importantly, we note that the parallax uncertainties of the M giants are all smaller than those of the Miras. This is shown in Fig. \ref{app_Mdwarf}, where the logarithmic value of the parallax uncertainty is plotted as a function of the logarithm of the distance (here simply taken as the inverse of the parallax). On average, the M giants have parallax uncertainties that are smaller by a factor of 3.5 than those of the Miras. As the M giants are more compact than the Miras and have smaller pressure scale heights, it is plausible that the larger parallax uncertainties of the Miras indeed result from their surface convection cells. We therefore conclude that our key assumption is correct.

\end{appendix}

\end{document}